%% file: main_condensed_cleanv1.tex
\documentclass[preprint,superscriptaddress,amsmath,amssymb,aps,onecolumn,floatfix,]{revtex4-2}

\usepackage{graphicx}
\usepackage{dcolumn}
\usepackage{bm}
\usepackage{amssymb}
\usepackage{amsmath}
\usepackage{parskip}
\usepackage{multirow}
\usepackage{xcolor}
\usepackage[sort&compress]{natbib}
\setcitestyle{super}
\usepackage{float}
\usepackage{units}
\usepackage[percent]{overpic}
\usepackage{braket} 
\usepackage[caption=false]{subfig} 
\usepackage{changepage}
\usepackage{amsmath}
\usepackage{xfrac}
\usepackage{appendix}
\usepackage{soul}
\usepackage[version=4]{mhchem}

\begin{document}


\title{Emergent Surface Kondo Flat Band Driven by Competing Interactions in a Topological Ferromagnet}


\author{Nazar Zaremba}\thanks{Nazar Zaremba, Susmita Changdar, Haojie Guo, Iñigo Robredo, and Daniel Lozano-G\'omez contributed equally to this work.}
\affiliation{Max Planck Institute for Chemical Physics of Solids,  01187 Dresden, Germany}
\author{Susmita Changdar}\thanks{Nazar Zaremba, Susmita Changdar, Haojie Guo, Iñigo Robredo, and Daniel Lozano-G\'omez contributed equally to this work.}
\affiliation{Leibniz Institute for Solid State and Materials Research, IFW Dresden, 01069 Dresden, Germany}
\author{Haojie Guo}\thanks{Nazar Zaremba, Susmita Changdar, Haojie Guo, Iñigo Robredo, and Daniel Lozano-G\'omez contributed equally to this work.}
\affiliation{Donostia International Physics Center, 20018 Donostia-San Sebastián, Spain}
\author{Iñigo Robredo}\thanks{Nazar Zaremba, Susmita Changdar, Haojie Guo, Iñigo Robredo, and Daniel Lozano-G\'omez contributed equally to this work.}
\affiliation{Luxembourg Institute of Science and Technology (LIST), L-4362 Esch/Alzette, Luxembourg}

\author{Daniel Lozano-G\'omez}\thanks{Nazar Zaremba, Susmita Changdar, Haojie Guo, Iñigo Robredo, and Daniel Lozano-G\'omez contributed equally to this work.}
\affiliation{Institut f\"ur Theoretische Physik and W\"urzburg-Dresden Cluster of Excellence ctd.qmat, Technische Universit\"at Dresden, 01062 Dresden, Germany}

\author{Andreas Leithe-Jasper}
\affiliation{Max Planck Institute for Chemical Physics of Solids, 01187 Dresden, Germany}

\author{Rui Lou}
\affiliation{Leibniz Institute for Solid State and Materials Research, IFW Dresden, 01069 Dresden, Germany}
\affiliation{Helmholtz-Zentrum Berlin f\"ur Materialien und Energie, 14109 Berlin, Germany}

\author{Yurii Prots}
\author{Mitja Krnel}
\author{Markus K\"onig}
\affiliation{Max Planck Institute for Chemical Physics of Solids, 01187 Dresden, Germany}
\author{Konstantin Semeniuk}
\affiliation{Institute for Quantum Materials and Technologies, Karlsruhe Institute of Technology, 76131 Karlsruhe, Germany}

\author{Berit H. Goodge}
\author{Yuri Grin}
\affiliation{Max Planck Institute for Chemical Physics of Solids, 01187 Dresden, Germany}

\author{Andrii Kuibarov}
\affiliation{Leibniz Institute for Solid State and Materials Research, IFW Dresden, 01069 Dresden, Germany}

\author{Oleksandr Suvorov}
\affiliation{Leibniz Institute for Solid State and Materials Research, IFW Dresden, 01069 Dresden, Germany}
\affiliation{Kyiv Academic University, 03142 Kyiv, Ukraine}

\author{Alexander Fedorov}
\affiliation{Leibniz Institute for Solid State and Materials Research, IFW Dresden, 01069 Dresden, Germany}
\affiliation{Helmholtz-Zentrum Berlin f\"ur Materialien und Energie, 14109 Berlin, Germany}

\author{Adam Pikul}
\affiliation{Institute of Low Temperature and Structure Research, Polish Academy of Sciences, 50-422 Wroclaw, Poland}
\affiliation{Centre for Advanced Materials and Smart Structures, Polish Academy of Sciences, 50-422 Wroclaw, Poland}

\author{Ivan Soldatov}
\author{Rudolf Sch\"afer}
\author{Bernd B\"uchner}
\affiliation{Leibniz Institute for Solid State and Materials Research, IFW Dresden, 01069 Dresden, Germany}

\author{Oksana Kvitnitskaya}
\affiliation{Leibniz Institute for Solid State and Materials Research, IFW Dresden, 01069 Dresden, Germany}
\affiliation{B. Verkin Institute for Low Temperature Physics and Engineering of the National Academy of Sciences of Ukraine, 61103 Kharkiv, Ukraine}

\author{Yurii Naidyuk}
\affiliation{B. Verkin Institute for Low Temperature Physics and Engineering of the National Academy of Sciences of Ukraine, 61103 Kharkiv, Ukraine}

\author{Louis Lamm}
\affiliation{Institute for Solid State and Materials Physics, Dresden University of Technology, 01062 Dresden, Germany}
\author{Meike Pfeiffer}
\affiliation{Max Planck Institute for Chemical Physics of Solids, 01187 Dresden, Germany}
\affiliation{Institute for Solid State and Materials Physics, Dresden University of Technology, 01062 Dresden, Germany}
\affiliation{Institute for Quantum Materials and Technologies, Karlsruhe Institute of Technology, 76131 Karlsruhe, Germany}

\author{Elena Hassinger}
\affiliation{Max Planck Institute for Chemical Physics of Solids, 01187 Dresden, Germany}
\affiliation{Institute for Quantum Materials and Technologies, Karlsruhe Institute of Technology, 76131 Karlsruhe, Germany}

\author{Matthias Vojta}
\affiliation{Institut f\"ur Theoretische Physik and W\"urzburg-Dresden Cluster of Excellence ctd.qmat, Technische Universit\"at Dresden, 01062 Dresden, Germany}

\author{Sergey Borisenko}
\affiliation{Leibniz Institute for Solid State and Materials Research, IFW Dresden, 01069 Dresden, Germany}

\author{Miguel M. Ugeda}
\affiliation{Donostia International Physics Center, 20018 Donostia-San Sebastián, Spain}
\affiliation{IKERBASQUE, Basque Foundation for Science, 48009 Bilbao, Spain}
\affiliation{Centro de Física de Materiales CSIC-UPV/EHU, 20018 Donostia-San Sebastián, Spain}

\author{Jeroen van den Brink$^{*}$}
\email{j.van.den.brink@ifw-dresden.de}
\affiliation{Leibniz Institute for Solid State and Materials Research, IFW Dresden, 01069 Dresden, Germany}
\affiliation{Institut f\"ur Theoretische Physik and W\"urzburg-Dresden Cluster of Excellence ctd.qmat, Technische Universit\"at Dresden, 01062 Dresden, Germany}

\author{Eteri Svanidze$^{*}$}
\email{svanidze@cpfs.mpg.de}
\affiliation{Max Planck Institute for Chemical Physics of Solids, 01187 Dresden, Germany}

\author{Maia G. Vergniory$^{*}$}
\email{maia.vergniory@usherbrooke.ca}
\affiliation{Donostia International Physics Center, 20018 Donostia-San Sebastián, Spain}
\affiliation{D\'epartement de Physique et Institut Quantique, Universit\'e de Sherbrooke, Sherbrooke, J1K 2R1, Qu\'ebec, Canada.}
\affiliation{Regroupement Qu\'eb\'ecois sur les Mat\'eriaux de Pointe (RQMP), H3T 3J7 Qu\'ebec, Canada}

\date{\today}

\begin{abstract}

%
%
%
A central goal of modern condensed matter physics is to uncover new quantum states of matter arising from the intertwined effects of strong electron correlations, magnetism, and band topology. Heavy-fermion phases, generated by Kondo interactions, represent one of the most remarkable manifestations of electronic correlations, and topological heavy-fermion states have been identified in several non-magnetic materials. Yet, the consequences of their competition with magnetic order have remained largely unexplored. Here, we reveal a new phenomenon: the spontaneous spatial separation of correlated quantum phases. By showing that magnetism can drive distinct strongly correlated electronic states to coexist in different regions of a single material, our work establishes a previously unknown mechanism for organizing quantum matter and opens a new direction in the study of correlated topological systems.
Using \emph{bulk-sensitive} probes, we show that UAsS crystals are, in the bulk, metallic ferromagnets with only moderate correlation-driven band renormalizations.
First-principles calculations reveal a topological electronic structure hosting both nodal lines and Weyl points, pointing to a rich underlying topology. Angle-resolved photoemission spectroscopy (ARPES) measurements are consistent with these predictions, resolving the nodal lines and Weyl crossings. In striking contrast, \emph{surface-sensitive} ARPES and scanning tunneling microscopy/spectroscopy (STM/STS) measurements reveal a pronounced flat band pinned at the Fermi level, accompanied by a sharp resonance -- hallmarks of an emergent, strongly correlated Kondo state not captured by first-principles calculations.
The existence of correlated heavy fermions confined to the surface is consistent with model calculations that incorporate the reduced coordination of the uranium atoms and the associated weakening of ferromagnetic order at the surface.
These findings establish UAsS as a unique platform in which a novel surface-confined many-body state emerges,  
demonstrating that competing interactions can selectively reshape surface electronic states, and give rise to correlation-driven features, absent in the bulk. This opens new avenues for engineering emergent many-body phenomena at crystalline surfaces.

\end{abstract} 

\maketitle

The competition between different forces typically creates fertile ground for novel states of matter to emerge. This holds for classical systems, but particularly also for electrons in a solid, which constitute a deeply entangled and interacting many-body quantum state. 
In heavy-fermion systems, intermetallic materials that contain atoms with partly filled $4f$ or $5f$ electronic orbitals, one main driving force is the Kondo interaction between localized $f$-electrons and itinerant conduction electrons. 
%
This interaction favors the formation of a highly correlated, non-magnetic Kondo-screened state populated by heavy quasiparticles with effective masses far larger than those of free band electrons.
%
%
Magnetic interactions are the nemesis of these heavy fermions, driving the $f$-electron moments instead towards a much more conventional, magnetically ordered collective state. As a result, the coexistence of long-range magnetic order and Kondo screening is both uncommon and highly nontrivial, making materials that host both phenomena especially valuable for exploring
new correlated quantum states.
When topological electrons are added to this competition, the resulting landscape becomes even richer: topology can protect unconventional boundary states and topologically charged quasiparticles, providing new channels through which electronic correlations, magnetism, and topology intertwine. Understanding how these competing tendencies are reconciled remains one of the central challenges in the physics of correlated quantum matter, and the discovery of new materials in which strong correlations, magnetism, and topology coexist has become a major priority in the field.

While strong electronic correlations are generally known to coexist with non-trivial band topology, the subsequent emergence of Kondo physics has so far been reported in only a few bulk crystals, including SmB$_6$, YbB$_{12}$, and SmAsS \cite{Xu2014,Ohtsubo2019,Iraola2023,Chang2017,Sundermann2015,Baruselli2014,Hagiwara2016,Robredo2025}, as well as Weyl-Kondo semimetals \cite{Lai2018,Grefe2020,Dzsaber2017}, exemplified by Ce$_3$Pd$_3$Bi$_4$ \cite{Dzsaber2017,Dzsaber2021}. 
In this context, the antiferromagnet CeCo$_2$P$_2$ provides an interesting example in which different correlated orbitals simultaneously host heavy-fermion correlations and magnetic order \cite{Hu2025,Liu2024}. A bulk Kondo resonance develops within the antiferromagnetically ordered phase, and the low-temperature electronic structure exhibits a nodal-line semimetal state, arising from the hybridization between Ce-$4f$ and Co-$3d$ states. Another example is the van der Waals antiferromagnet UOTe, where a temperature-dependent U-$5f$ excitation, interpreted as evidence of Kondo screening, emerges within the antiferromagnetic phase and coexists with a bulk Dirac semimetal state \cite{Broyles2025}. In both systems, the correlated and topological phenomena are bulk properties and occur in antiferromagnetic backgrounds. Moreover, the experimentally observed topological features correspond to symmetry-protected nodal-line or Dirac semimetal states, whose band crossings do not carry a net topological charge. As a result, the topology is encoded in symmetry-enforced degeneracies rather than in charged topological quasiparticles such as Weyl fermions.

In this work, we present UAsS \cite{Robredo2024,TQC,MTQC}, a rare example of a topological ferromagnet based on $5f$ electrons—a regime in which the coexistence of topology, ferromagnetism, and Kondo physics is particularly challenging. In conventional Kondo lattices, antiferromagnetic order, driven by the oscillatory nature of the RKKY interaction, can coexist with Kondo screening and give rise to heavy-fermion antiferromagnets. By contrast, ferromagnetism is much rarer and remains comparatively poorly understood. Experimental studies of several Ce-based compounds indicate that the competition between Kondo screening and ferromagnetic order is often resolved through the formation of modulated magnetic phases, thereby destabilizing uniform ferromagnetism \cite{PhysRevB.98.195119}. This competition is expected to be even more severe in $5f$ systems. Unlike the more localized $4f$ electrons found in conventional Kondo materials, $5f$ electrons are significantly more itinerant and hybridize more strongly with their environment, blurring the distinction between localized and itinerant degrees of freedom. As a result, the formation of robust local moments—and hence Kondo screening itself—is far less straightforward.
Here, we show that this competition is resolved in a novel elegant manner. Rather than one phenomenon suppressing the other, the two become spatially separated: ferromagnetism and topological electronic states dominate the bulk, while Kondo screening emerges selectively at the surface. This correlation-driven bulk–surface differentiation provides a previously unexplored mechanism for accommodating competing quantum phenomena within a single material. More broadly, it establishes a platform in which distinct electronic phases coexist in real space and can, in principle, be manipulated independently, opening new opportunities for the design of correlated and topological quantum functionalities.
Strong and complementary experimental evidence from ARPES, STM/STS, and bulk-sensitive probes establishes both the weakly correlated ferromagnetic character of the bulk and the emergence of the correlated surface state. As a result, an unprecedented surface Kondo state develops at U-terminated surfaces, characterized by a remarkably strong Kondo resonance pinned at the Fermi level, with a surface Kondo temperature of $T_{\rm K}\sim52$ K.

Uranium-based compounds are in general fertile platforms for emergent quantum phenomena because their $5f$ electrons occupy an intermediate regime between itinerancy and local-moment behavior, giving rise to a wide range of competing electronic instabilities and collective states, including unconventional magnetism, heavy-fermion behavior, quantum criticality, hidden and multipolar orders, and unconventional superconductivity \cite{Moriya1985,Lee2018,Christovam2024,Amorese2020,Marino2024,Zwicknagl2003,Aoki2001,Gill2011,Stewart1984,Wirth2016,Brando2016}. At the same time, a growing number of uranium compounds have been proposed to host topological electronic structures and associated surface states \cite{Christiansen2025,Choi2024,Broyles2025,Broyles2025b,Siddiquee2023,Goswami2013,Schemm2015,Goswami2015,Yanase2017,Tsutsumi2013,Giannakis2019,Chen2019}. The surface-confined Kondo state that we observe in UAsS suggests that the spatial separation of correlations uncovered here may be a more general phenomenon in topological uranium materials, opening a route toward the coexistence and interplay of surface Kondo physics with a broad range of competing many-body states.

\begin{figure}[!t]
\includegraphics[width=\textwidth]{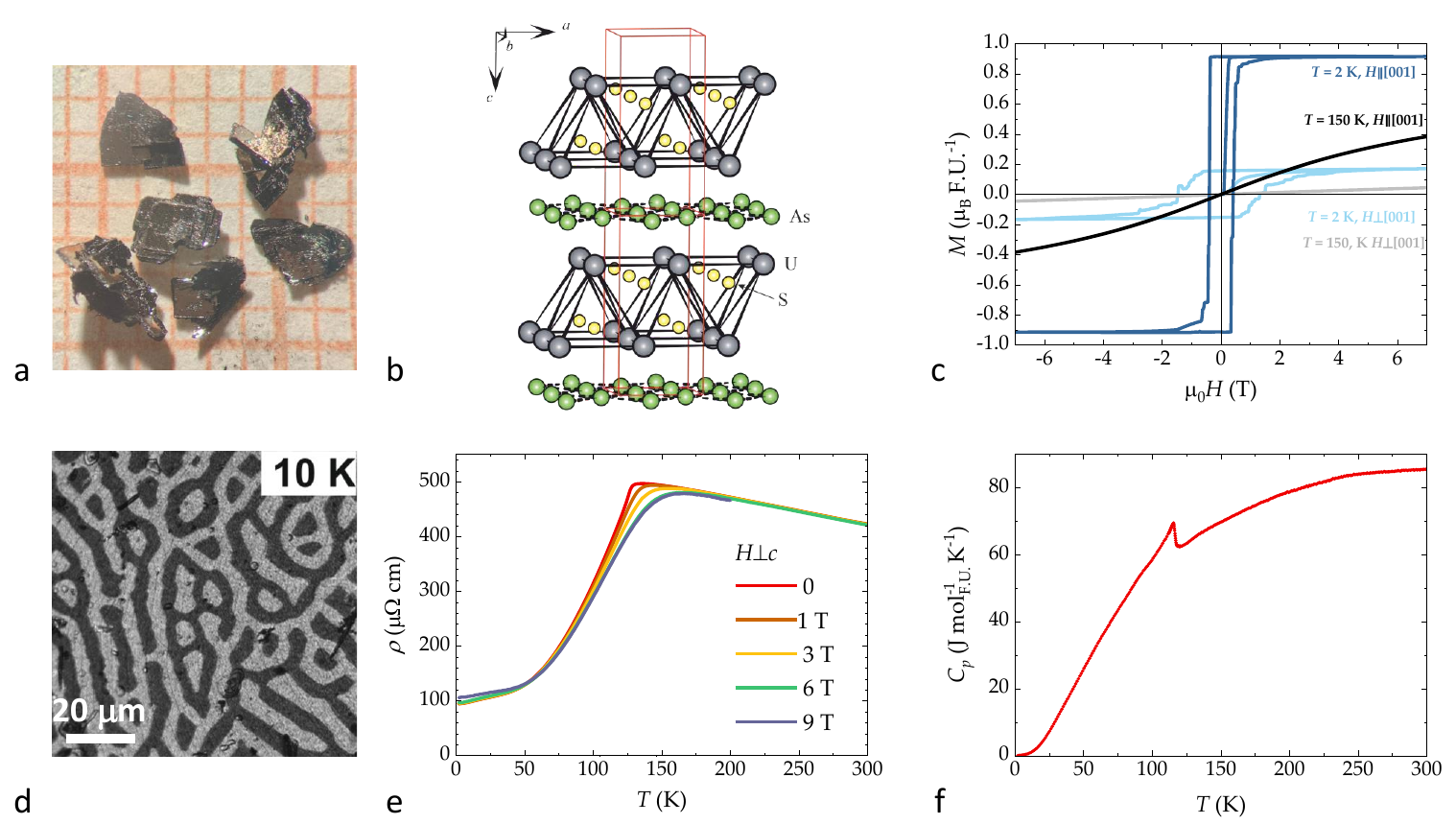}
\caption{\footnotesize \setlength{\baselineskip}{0.8\baselineskip} 
\textbf{Crystal structure and bulk ferromagnetism of UAsS.} (a) Single crystals of UAsS typically have plate-like morphology with the $c$-axis being perpendicular to the largest facet. (b) Within the lattice of UAsS, two termination planes are possible -- either along the As or along the U/S layer. (c) Bulk ferromagnetic order occurs below $T_\mathrm{C} = 126$ K, with high level of anisotropy. (d) Magnetic domains observed on the basal plane of the UAsS single crystal in ferromagnetic state. (e, f) Bulk ferromagnetic order is marked by a corresponding features in the resistivity, as well as specific heat data.}
\label{fig:Structure}
\end{figure}

\noindent {\bf Structure and bulk thermodynamic properties}\\
We have grown UAsS in large, millimeter sized platelet-like crystals, with the $c$-axis being perpendicular to the plane of the crystals (see Fig.~\ref{fig:Structure}a and \ref{Synthesis}). It crystallizes in the tetragonal PbClF (ZrSiS) structure type (space group $P4/nmm$) -- see Fig.~\ref{fig:Structure}b. Sulfur atoms are coordinated by 4+1 U atoms, forming square pyramids. These pyramids are condensed via edges into layers, which are perpendicular to [001]. These layers alternate with planar As layers as shown in Fig.~\ref{fig:Structure}b, with crystalline stacking faults observed by scanning transmission electron microscopy (\ref{STEM}). The layered nature of the structure supports cleavage of the crystal along the $ab$-plane, also identified by chemical bonding analysis as the most probably cleavage planes. Furthermore, the most energetically favorable termination plane is the U/S one (see Fig.~\ref{fig:Structure}b and \ref{Synthesis}).
The U--U interatomic distances are 3.877(1) \AA, which are longer than the distances in elemental uranium (2.75-3.43 \AA) \cite{Donohue1974}. According to the empirical Hill criterion \cite{Hill1970}, such a large separation is likely to yield a magnetic ground state. Indeed, we observe, consistent with literature \cite{Hulliger1968,Bazan1972, Wojakowski1972}, that in our crystals bulk ferromagnetism sets in below $126$ K (see Fig.~\ref{fig:Structure}c-f, and Fig.~\ref{fig:BulkFM}, as well as \ref{bulk}, \ref{Kerr}, and \ref{PCS}). Unlike in the case of UAs$_2$ \cite{li2026}, application of moderate hydrostatic pressure increases the value of $T_C$ in UAsS (see \ref{Pressure}).

The effective magnetic moments extracted from a Curie-Weiss fit are 2.2 $\mu_B$ per U for $H$$\parallel$$[001]$ and 0.6 $\mu_B$ per U for $H$$\perp$$[001]$, consistent with the ordered magnetic moment $\mu_{ord} = 1.24$ $\mu_B$ measured by neutron scattering \cite{Zygmunt1974}. The entrance into ferromagnetic state is marked by a corresponding anomaly in specific heat and electrical resistivity data -- see Fig.~\ref{fig:Structure}e and f. 
As reported previously \cite{Wojakowski1972, Wojakowski1987, Henkie1998}, 
bulk electrical resistivity of UAsS shows a large anomaly associated with magnetic order, indicative of a strong coupling between magnetic and electronic degrees of freedom.

The specific heat of UAsS exhibits typical features of intermetallic compounds, with a phonon background corresponding to a Debye temperature of the order of room temperature and a typical $\Lambda$ anomaly at the ferromagnetic ordering temperature. A standard Sommerfeld–Debye analysis in the low-temperature regime yields a small electronic term ($\gamma = 0.022$ J K$^{-2}$mol$^{-1}$), consistent with the presence of magnetic order, weak Kondo screening of the magnetic moments and the absence of any heavy-fermion physics in bulk UAsS. 
\begin{figure*}[htbp!]
    \centering
    \includegraphics[width=\textwidth]{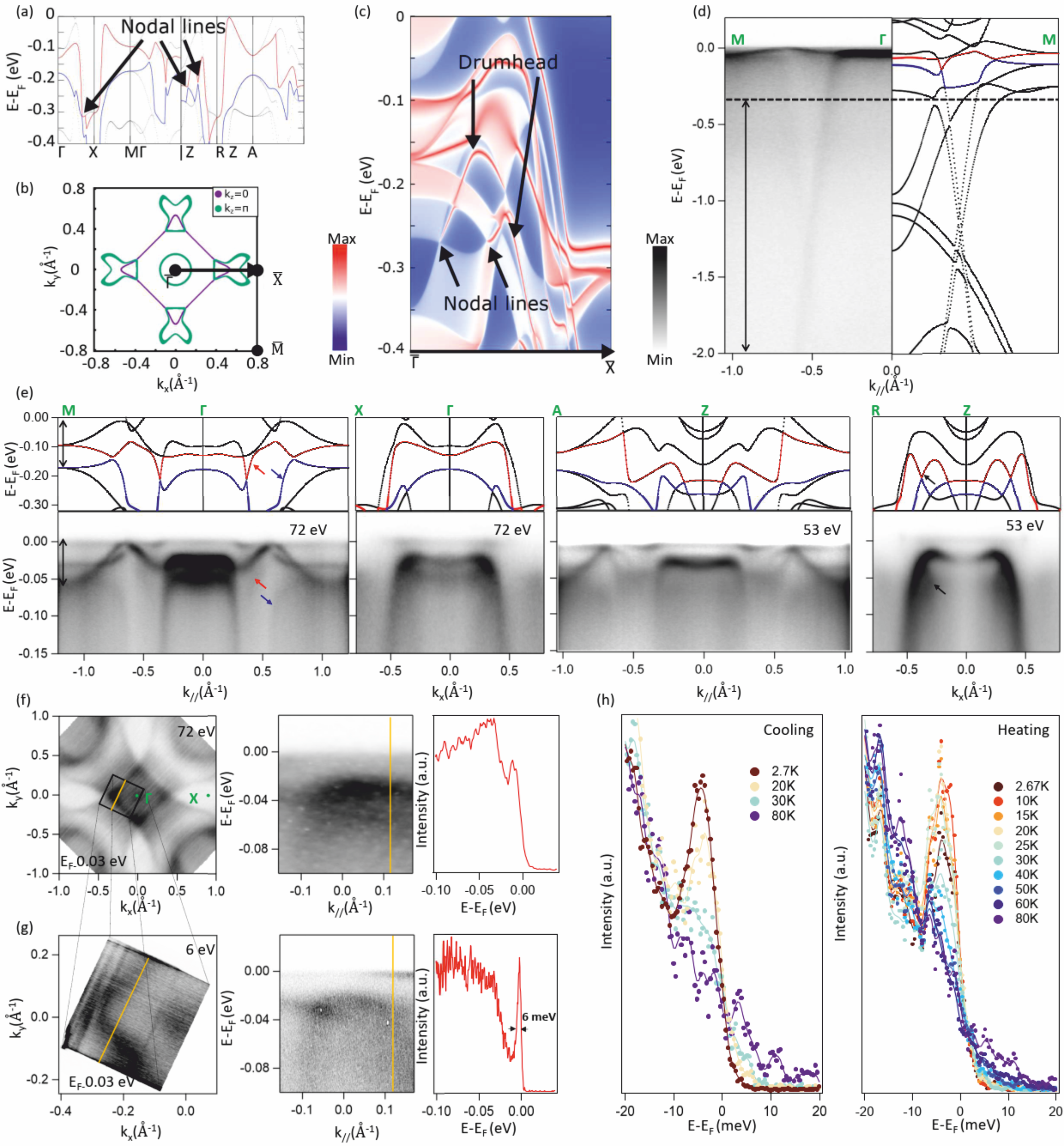}
    \caption{\footnotesize \setlength{\baselineskip}{0.8\baselineskip} \textbf{Electronic structure of UAsS, from electronic structure calculations and ARPES measurements.}
    (a) Zoomed-in band structure below the Fermi level. In red/blue, the electronic bands whose crossings form nodal lines (pointed with arrows). (b) Nodal lines at $k_z=0,\pi$ planes. The arrow indicates the direction in momentum space for the surface spectrum. (c) (001) surface spectrum. We can see drumhead states stemming from the surface projection of the nodal lines. Notice how drumhead states exist in the regions where the projection of the nodal lines does not overlap. (d) Long range EDM along ($\Gamma-M$) compared with calculated bands. (e) Comparison between DFT bulk band structure calculations (top panels) and experimental EDMs (bottom panels) along $\Gamma-M$, $\Gamma-X$ ($h\nu = 72$ eV) and $Z-A$, $Z-R$ ($h\nu = 53$ eV), highlighting the strong band renormalization near E$_F$ and nodal line features (red/blue arrows and black marker). (f) Constant energy contour at $E_F-0.3$ eV measured with $h\nu = 72$ eV, corresponding EDM, and EDC taken along the yellow line. (g) Constant energy contour at $E_F-0.3$ eV measured with $h\nu = 5.9$ eV, corresponding EDM, and the EDC showing a sharp peak near $E_F$. (h) Temperature dependence of the EDC from the flat band measured with $h\nu = 5.9$ eV during cooling and heating.}
    \label{fig:ARPES_2}
\end{figure*}

\noindent {\bf Electronic structure from ARPES and DFT}\\
Next, we investigated the electronic structure of UAsS by Angle-Resolved Photoemission Spectroscopy (ARPES) with synchrotron radiation and laboratory-based laser ($h\nu = 5.9$ eV), see \ref{Arpes}. The out-of-plane Fermi surface (FS) map reveals the overall symmetry of the electronic states and the high-symmetry points at 72 eV and 53 eV photon energies (Fig.~\ref{Figure S1:ARPES}b). To confirm the locations of $\Gamma$ and $Z$ points, FS maps taken with $h\nu$=72 eV and $h\nu$=53 eV (Fig.~\ref{Figure S1:ARPES}c) are compared with calculated FS maps at $k_z$=0 and $k_z$=$\pi$ (Fig.~\ref{Figure S1:ARPES}d), which confirms that $72$ eV corresponds to the $k_z = 0$ plane and $53$ eV to the $k_z = \pi$ plane. 

Inspection of the near-Fermi region with first principles electronic structure calculations (Fig.~\ref{fig:ARPES_2}a) reveals band crossings with distinct orbital character (see \ref{DFT}), giving rise to the presence of topological nodal lines, as indicated by the arrows. The momentum-space distribution of these nodal lines is shown in Fig.~\ref{fig:ARPES_2}b, where they form extended loops in both the $k_z=0$ and $k_z=\pi$ planes. Such nodal lines are expected to present drumhead surface states on the (001) surface within regions of momentum space enclosed by the projection of a single nodal line \cite{Fang2016,Robredo2022}. Consistent with this expectation, Fig.~\ref{fig:ARPES_2}c displays the drumhead-like surface states that emerge from the projected nodal lines. These states are clearly resolved in regions where the projections of different nodal lines do not overlap, confirming the anticipated bulk-boundary correspondence. Time-reversal symmetry breaking allows the presence of Weyl nodes and indeed we have identified 3 families of Weyl nodes very close to the Fermi level, both of type I and type II~\cite{Soluyanov2015} (see \ref{DFT}).

To probe the calculated topological features experimentally, detailed Energy Distribution Maps (EDMs) were collected along key high-symmetry directions, as shown in the bottom panel of Fig.~\ref{fig:ARPES_2}e. Specifically, EDMs along $\Gamma-M$ and $\Gamma-X$ were measured with $h\nu=$ $72$ eV, while those along $Z-A$ and $Z-R$ were collected with $h\nu=53$ eV. These experimental EDMs are directly compared with their corresponding DFT bulk band structure calculations (top panels of Fig.~\ref{fig:ARPES_2}e). A striking observation is the band renormalization by a factor of three for bands close to the Fermi level ($E_F$) (Fig.~\ref{fig:ARPES_2}d). This renormalization is a signature of electron-electron interactions and is associated to the U-$5f$ orbital character to these states (Fig.~\ref{fig:fig_dft_2}a). The pronounced renormalization makes it challenging to resolve the topological nodal lines experimentally, but the EDMs do reveal signatures of the nodal lines: bands approaching the nodal points are visible along $\Gamma$–$M$ (highlighted by red and blue arrows) and nodal point along $Z$–$R$ (marked in black). Moving approximately 0.3 eV below $E_F$, the bands mainly have As-$4p$ character (Fig.~\ref{fig:fig_dft_2}a) and the band renormalization factor reduces to 1.

Interestingly, ARPES also clearly reveals the presence of a $k_z$-independent flat band at the Fermi level, in all EDMs (Fig.~\ref{fig:ARPES_2}e). This flat feature is not captured by bulk DFT calculations and directly points towards an emergent Kondo screening of local moments. To investigate the Kondo-resonance character of this feature in more detail, we performed high-resolution Laser ARPES measurements using a 5.9 eV laser source. Fig.~\ref{fig:ARPES_2}g shows a constant energy contour at $E_F - 0.3$ eV (left panel), and the corresponding EDM along the yellow line reveals a sharp, flat band near the Fermi level. The full width at half maximum (FWHM) of this band is approximately 6 meV, which showcases the high-resolution of the set-up. 
The fact that the flat band is so very sharp in energy evidences directly that we are dealing with a surface state.
To indicate the region of the Brillouin zone (BZ) probed by the Laser ARPES measurements, we present the corresponding synchrotron data in Fig.~\ref{fig:ARPES_2}f. We chose the constant energy contour at 0.3 eV below the Fermi level because the lack of bands near the $\Gamma$ point at $E_F$ makes it difficult to clearly determine the region of the BZ covered by the Laser measurements. Next, we performed temperature-dependent Laser ARPES measurements (Fig.~\ref{fig:ARPES_2}h), which show that the EDC peak associated with this flat band consistently disappears above $\sim$ 50 K and reappears upon cooling.

At around $\sim$ 50 K, which is firmly below the ferromagnetic ordering temperature of UAsS at 126 K, there are no signatures of any anomaly in the bulk thermodynamic, magnetic and transport properties. This temperature dependence combined with its $k_z$-independent character, the location at the Fermi energy and its absence from band structure calculations, points towards the presence of a many-body Kondo-resonance state at the surface.

\begin{figure}
    \centering
    \includegraphics[width=1\linewidth]{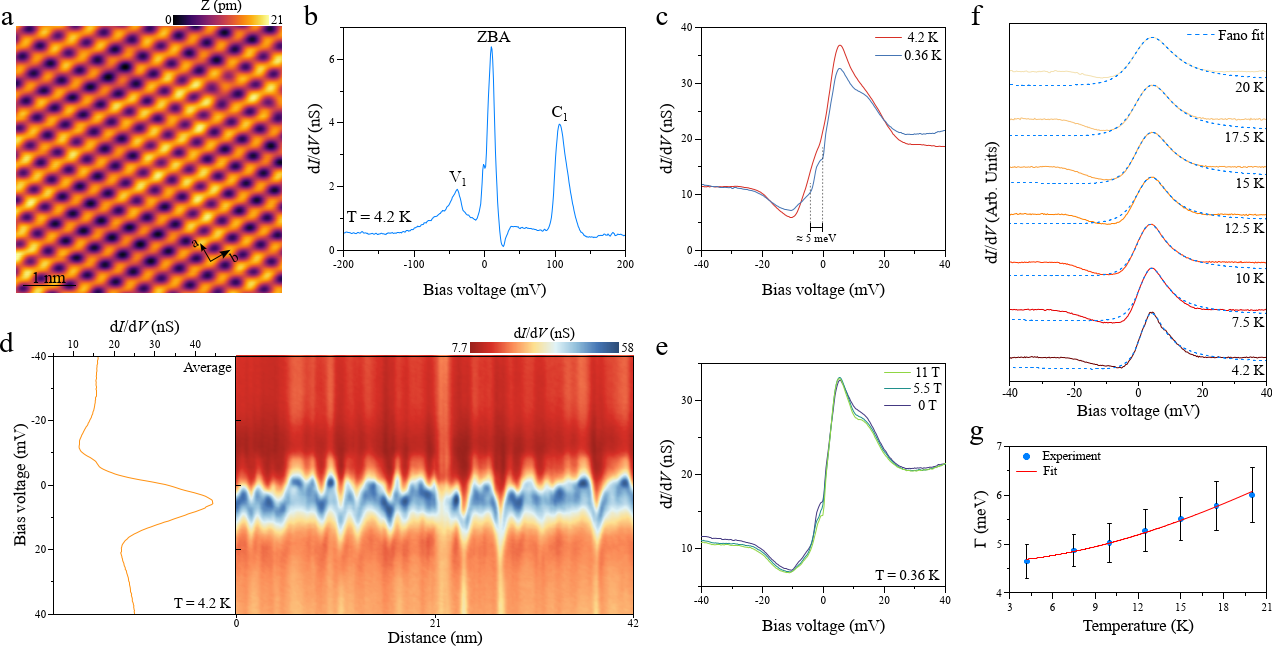}
    \caption{\footnotesize \setlength{\baselineskip}{0.8\baselineskip} \textbf{Local tunneling spectroscopy characterization of UAsS.} (a) Atomically resolved STM topography image of the U-terminated surface. (b) Representative d$I$/d$V$ spectrum of the low-lying electronic structure of uranium surface. (c) Spatially averaged d$I$/d$V$ spectra probing the ZBA at U-planes near $E_{F}$. (d) One-dimensional d$I$/d$V$($r$,$V$) map along 42 nm revealing the spatial variations of the ZBA. (e) Out-of-plane magnetic-field ($H_{\perp}$) dependence of the ZBA. (f) Set of d$I$/d$V$ spectra recorded in the same sample location showing the temperature evolution of the ZBA and the corresponding fits to a Fano line shape (dashed blue lines). (g) Extracted values of the intrinsic width of the ZBA from each Fano fit in (f). The fitted red solid line represents the expected intrinsic Kondo peak width ($\Gamma$) evolution with temperature (see main text). Acquisition parameters: (a) $V_{\mathrm{s}} = 70$ mV, $I_{\mathrm{t}} = 0.3$ nA, $T = 4.2$ K. (b) $V_{\mathrm{a.c.}} = 1$ mV, $T = 4.2$ K. (c)-(f) $V_{\mathrm{a.c.}} = 0.3$ mV.}
    \label{fig:STM}
\end{figure}

\noindent {\bf STM/STS: zero bias anomaly and surface Kondo resonance}\\
To further experimentally probe the low-energy quasiparticle excitations at UAsS surfaces, we performed atomic-scale characterization using scanning tunneling microscopy/spectroscopy (STM/STS) at low temperatures (4.2 K and 0.36 K) and high magnetic fields. Upon the cleavage of the UAsS crystal (\ref{STM}), we observe two types of exposed surfaces with square atomic arrangements with in-plane periodicity $\simeq 3.85 \pm 0.05$ \AA~ (Fig.~\ref{fig:STM}a) and $\simeq 2.74 \pm 0.05$ \AA~ (Fig.~\ref{Figure S1:STM}). While the latter corresponds unequivocally to the As-plane, for the former both the U- ans S- terminations are, in principle, compatible. However, our electron localizability indicator calculations (ELI-D) suggest that cleavage is expected to occur along planes associated with the S- and As-centered lone-pair regions (\ref{Synthesis}). Since cleavage between the U-S and As layers yields complementary As- and U-terminated surfaces, we assign the periodicity $\simeq 3.85$ \AA~ to the U-terminated plane (\ref{STM}). Furthermore, comparison between atomically resolved STM images of the S-, As- and U-termination planes and corresponding DFT calculations 
enable us to rule out the visualization of the S-plane \cite{Guo2026}.

Next, we determined the electronic structure using STM spectroscopy. Differential tunneling conductance (d$I$/d$V$) measurements (Fig.~\ref{fig:STM}b) show that the low-energy electronic structure of the U-plane is dominated by three sharp resonances. Below and above $E_{F}$, we find two peaks, labeled as $V_{1}$ and $C_{1}$, respectively, which we assign a band-structure origin. In the negative bias (occupied states) region, the most pronounced feature is the peak labeled $V_{1}$ at $V_{S} = -30$ mV. Its energy coincides with the binding energy of the 5$f$-derived flat band at $\Gamma$ in the BZ, as seen in the ARPES spectra in Fig.~\ref{fig:ARPES_2} and Fig.~\ref{Figure S2:STM} (\ref{STM}). For positive bias (filled states), we observe the peak $C_{1}$ at $V_{S} = 120$ mV. According to our DFT calculations, this feature can be attributed to U-derived bands that flatten near $\Gamma$ at this energy (see \ref{DFT}).

In addition to these two band-structure features, a prominent peak centered around $E_{F}$ is present in the U-terminated surface, which we refer to as a zero-bias anomaly (ZBA) as it is not expected from \textit{ab}-initio calculations \ref{fig:ARPES_2}. Interestingly, we do not observe this feature at As-terminated surfaces (\ref{Arpes}). High-resolution d$I$/d$V$ spectra recorded at 4.2 K (Fig.~\ref{fig:STM}c) show that the ZBA peak has an asymmetric shape with an energy width of $25 \pm 5$ meV, and where the peak maximum is located few meV above $E_{F}$. Notably, part of the ZBA peak crosses $E_{F}$ as a decaying tail, thus leaving a trace of density of states below $E_{F}$. The d$I$/d$V$ spectra at lower temperatures of 0.36 K  (blue curve in Fig.\ref{fig:STM}c), are very similar and show that the tail of the ZBA below $E_{F}$ has an energetic width of $\approx 5 \pm 3$ meV, which is consistent with the $k_z$-independent flat band observed in ARPES EDMs (Fig.~\ref{fig:ARPES_2}g and h). We note from the STS data that the ZBA shows a spatial inhomogeneity in terms of peak intensity, position, and width across different sample locations (Fig.\ref{fig:STM}d), which is in-line with it being a quasiparticle excitation peak. 

To gain knowledge about this zero-bias spectroscopic feature, we conducted out-of-plane (parallel to the $c$-axis) magnetic field- and temperature-dependent STS measurements. Fig.\ref{fig:STM}e shows the evolution of the ZBA peak with $H_{\perp}$ up to 11 T. As seen, the manifest insensitivity of the ZBA to the field magnitude and orientation (\ref{STM}) suggests that likely it is not the fingerprint of a magnetic phase at the U-terminated surface, as proved for the bulk counterpart. 

The ZBA peak is well described by an asymmetric Fano-line shape, closely resembling zero-bias features reported in STM/STS studies of Kondo lattices in heavy-fermion materials \cite{Seiro2018,Giannakis2019,Jiao2020,Pirie2023,Giannakis2022,Ernst2011} as well as single-impurity Kondo systems\cite{Nagaoka2002,Ruan2021,Vao2021,Wan2023,Kruger2005}. In STS, such a Fano-line shape is commonly interpreted as the spectroscopic fingerprint of a Kondo resonance in the differential conductance, reflecting interference between tunneling into a narrow many-body resonance and into broader itinerant states. In our case, although bulk UAsS exhibits ferromagnetic order, Kondo screening may still develop at the exposed U-terminated surface, where the reduced atomic coordination weakens the exchange field relative to the bulk and can make ferromagnetic order no longer the favored local ground state. This may enable the emergence of a surface Kondo state, whose fingerprint in our d$I$/d$V$ spectra is the observed quasiparticle resonance peak.

To further test this interpretation of the ZBA, we tracked its temperature evolution (solid lines in Fig.~\ref{fig:STM}f) and found that the peak gradually broadens with increasing temperature until it ultimately vanishes.
This thermal evolution is well captured by a the \ref{STM}), from which we extract the temperature dependence of the intrinsic Kondo resonance width, $\Gamma$ (Fig.~\ref{fig:STM}g. In particular, for $T \ll T_K$, the intrinsic Kondo width\cite{Nagaoka2002,Ruan2021,Otte2008,Ternes2008,Zhang2013,Ternes2017,Gruber2018,Aynajian2010} is given by
\[
2\Gamma = \sqrt{(\alpha k_B T)^2 + (2 k_B T_K)^2},
\]
where $\alpha$ is a dimensionless constant that equals $2\pi$ within Fermi-liquid theory, and $T_K$ is the characteristic Kondo temperature. Fitting the temperature-dependent STS data to this expression yields $T_K = 52 \pm 5$ K and $\alpha = 4.8 \pm 0.5$. This value of $T_K$ agrees remarkably well with the temperature at which the Kondo-resonance flat band observed in ARPES disappears.

\begin{figure*}
    \centering
    \includegraphics[width=\textwidth]{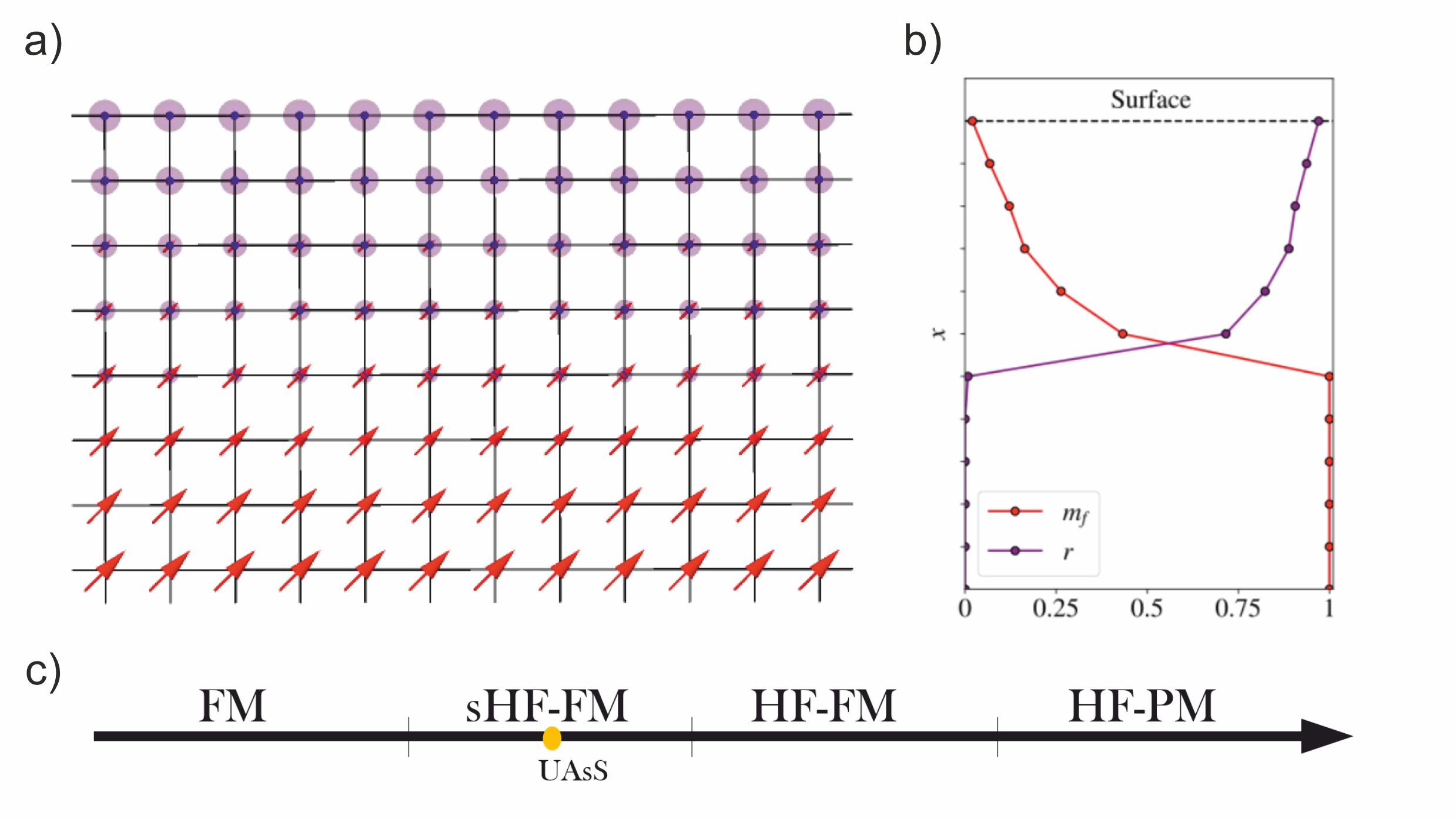}
    \caption{{Surface-induced Kondo screening.} (a) Illustration of the spatial evolution of magnetization and Kondo screening near the sample surface; the size of the red arrows (purple spheres) indicates the magnitude of the magnetization (screening) at the corresponding site. (b) Spatial evolution of the magnetization $m$ and screening, measured by the hybridization $r$, close to the boundary obtained via MFT. The MFT parameters are normalized to their saturation value. (c) Schematic phase diagram of the Kondo-Heisenberg model as a function of the Kondo coupling $J_K$, evolving from a ferromagnetic phase (FM) to a heavy-fermion paramagnet (HF-PM) via two intermediate phases, the surface heavy-fermion Ferromagnet (sHF-FM) and the regular (bulk) heavy-fermion Ferromagnet (HF-FM).}
    \label{Figure_6:MFT}
\end{figure*}


\noindent {\bf Kondo model and surface-induced screening}\\
Having experimentally identified the fingerprints of a surface Kondo resonance in both real and momentum space on U terminations in bulk ferromagnetic UAsS, we now develop a qualitative theoretical understanding of this complex microscopic interacting electronic structure and phenomenology on a model level. To this end we employ a standard Kondo-Heisenberg lattice model that captures the basic competing interactions -- Kondo screening and ferromagnetic ordering of the local moments. The fundamental question is whether, in the presence of bulk ferromagnetic order, a lattice boundary/surface may locally shift the balance toward screening. Numerically solving the model using a parton mean-field approximation on finite lattices with open boundary conditions (for details, see \ref{MFT}) we find a robust interaction window where, indeed, Kondo screening exists only at the boundary while ferromagnetic order which prevails in the bulk is dramatically weakened at the surface, see Fig.~\ref{Figure_6:MFT}. On the level of the model the driving force of having lower lattice coordination at the surface already suffices to produce a Kondo boundary state. The surface atoms having fewer neighbors reduces their exchange field compared to the bulk, and the resulting weakening of magnetic order tips the balance, giving way to a strongly enhanced Kondo screening, which provides the key mechanism for the formation of the very narrow electronic resonance observed at the surface of UAsS.

\noindent {\bf Outlook}\\
%
The realization of a correlated quantum state in which magnetism, non-trivial band topology, and Kondo correlations manifest in a spatially differentiated manner -- as we have demonstrated here in UAsS -- is highly unconventional. Kondo physics is well established in $4f$ systems, its realization in uranium-based $5f$ materials is counteracted by strong hybridization and electron delocalization, which in turn creates a rich platform for non-trivial band topologies. In UAsS the reduced lattice coordination at the surface weakens magnetic order and locally restores Kondo screening. It stabilizes a strongly correlated state with Kondo-screened moments confined to the surface, while the bulk remains a ferromagnetic metal.
This sharply contrasts with known systems such as CeCo$_2$P$_2$, where correlations, magnetism, and topology all coexist in the bulk. Instead, UAsS represents a new class of magnetic Kondo systems in which these interactions enable the emergence of surface phenomena. 
It establishes UAsS as a platform in which topology, magnetism, and correlations are disentangled in real space. More broadly, the phenomenology observed in UAsS guides a route toward discovering, tuning and engineering Kondo-driven surface resonances in topological metallic magnets that are otherwise weakly interacting, expanding the landscape of correlated topological matter.

\noindent {\bf Acknowledgements}\\
M.G.V received financial support from the Canada Excellence Research Chairs Program for Topological Quantum Matter. M.G.V and I.R. thank support to the grant PID2022-142008NB-I00 funded by MICIU/AEI/10.13039/50110001 and FEDER, UE, and the Ministry for Digital Transformation and of Civil Service of the Spanish Government through the QUANTUM ENIA project call -- Quantum Spain project, and by the European Union through the Recovery, Transformation and Resilience Plan - NextGenerationEU within the framework of the Digital Spain 2026 Agenda. M.M.U. acknowledges support by the ERC Starting grant LINKSPM (Grant $\#758558$) and by the grant PID2023-153277NB-I00 funded by MCIN/AEI/10.13039/501100011033. L.L., M.P., K.S., E.H., R. L., A. F., B.B., D.L.G., M.V., M.G.V., and J.v.d.B. acknowledge funding by the DFG through SFB 1143 (project ID 247310070) and the W\"urzburg-Dresden Cluster of Excellence on Complexity, Topology and Dynamics in Quantum Matter -- ctd.qmat (EXC 2147, project ID 390858490). L.L., M.P., K.S., and E.H. are supported by the ERC grant (Ixtreme, GA 101125759). O.K., Y.N., and N.Z. acknowledge the funding of Alexander von Humboldt Foundation (Philipp Schwartz-Initiative and Research Group Linkage Programme) and the support by the National Academy of Sciences of Ukraine under Project $\Phi 19-5$. A.P. acknowledges support from the Polish-U.S. Fulbright Commission through the Fulbright Senior Award 2025–2026, and the hospitality of the Idaho National Laboratory, including the INL Glenn T. Seaborg Institute and the Center for Quantum Actinide Science and Technology (C-QAST). S.C. and B.B. acknowledges the support by BMFTR funding through project 01DK240008 (GU-QuMat). O.S., B.B., and S.B. acknowledge the support of BMFTR through project “Instant micro-ARPES for in-operando tuning of material and device properties” (Project No. 05K2022-ioARPES). A.K. acknowledges the support of Deutsche Forschungsgemeinschaft through project no. 555830981. E.S. is grateful for the support of the Christiane N\"usslein-Volhard-Stiftung. E.S., N.Z., and M.K. acknowledge the support of the Boehringer Ingelheim Plus 3 Program. 

\bibliographystyle{naturemag}
\bibliography{References}

\input{Supplemental}

\end{document}

%% file: Supplemental.tex
\title{Kondo surface state emerging from competing magnetic and Kondo interactions in a topological metal UAsS}

\renewcommand{\thetable}{S\arabic{table}}
\renewcommand{\thefigure}{S\arabic{figure}}
\setcounter{figure}{0}
\setcounter{table}{0}
\renewcommand{\figurename}{FIG.}
\renewcommand{\tablename}{TABLE}

\renewcommand{\thesection}{Supplementary Note \Alph{section}}

\maketitle
\begin{center}
\textbf{\large{Supporting Information}}
\end{center}
\vspace{-10pt}
\section{Synthesis and crystal structure analysis}\label{subsec:synthecrystalstruct} \label{Synthesis}

All sample preparation and handling were performed in a specialized laboratory\cite{LeitheJasper2006}, equipped with an argon-filled glove box system (MBraun, p(H$_2$O/O$_2) < 0.1$ ppm). Single crystals of UAsS were grown by a chemical vapor transport reaction \cite{Schaefer1968, Binnewies2013} using arsenic triiodide (AsI$_3$) as the transport agent. For this purpose, natural U (wire cleaned using HNO$_3$, Goodfellow, 99.98\%), As (chunks, Alfa Aesar, Puratronic, 99.9999\%), and S (pieces, ChemPur, 99.999\%) in a temperature gradient from 980 $^\circ$C (source) to 890 $^\circ$C (sink), with a transport additive concentration of 10 mg/cm$^3$ AsI$_3$ (pieces, ThermoScientific, 99.999\%) were used. The resulting samples, consisting of shiny, gray, plate-shaped crystals of UAsS, were cleaned using acetone in a fume hood. The crystals of UAsS are stable in air for several weeks -- no noticeable degradation was observed either in bulk or on the microscale.

Powder X-ray diffraction was performed with a Huber G670 image-plate Guinier camera equipped with a Ge monochromator (CuK$\alpha_1$ radiation, $\lambda$ = 1.54056 \AA). Phase identification was done using the Match 3! software\cite{Putz2023}. For the single-crystal experiment, small pieces with dimensions about 10$\times$25$\times$25 $\mu$m were chosen to reduce absorption effect (79.950 mm$^{-1}$). The diffraction data were collected using a Rigaku AFC7 diffractometer equipped with a Saturn 724+ CCD detector. The data reduction was performed by using CrysAlis$^{PRO}$SM software\cite{CrystalClear2011}. An empirical absorption correction was performed by a multi-scan routine\cite{Blessing1995}. The WinGX suite of programs\cite{Farrugia2012} was used for structure solution, refinement, and data analysis. For the final runs, lattice parameters obtained from the powder diffraction data using individual peak positions (extracted by profile fitting and corrected with LaB$_6$ as an internal standard using the WinCSD software package\cite{Akselrud2014}) were used. The complete crystallographic information, atomic coordinates with equivalent displacement parameters and interatomic distances are listed in the Supplementary Tables \ref{tab:crystaldata}, \ref{tab:coordinates}, and \ref{tab:distances}, respectively. The resultant cell parameter values (see Supplementary Table \ref{tab:latpar}) are very similar to the values reported in previous works, signaling a fully ordered structure and excluding a homogeneity range. 


\setlength{\tabcolsep}{12pt}
\begin{table}
\begin{footnotesize}
\centering
\caption{{Crystallographic data and structure refinement of UAsS (PbClF-type)}. {\footnote{The cif file has been deposited at the Cambridge Crystallographic Data Center (CCDC number 2551015) and contains the supplementary crystallographic data for this paper. These data can be obtained free of charge via \href{https://www.ccdc.cam.ac.uk/data_request/cif}{website}, by \href{mailto:data_request@ccdc.cam.ac.uk}{email}, or by contacting the Cambridge Crystallographic Data Centre, 12 Union Road, Cambridge CB2 1EZ, UK; fax: +44 1223 336033.}}}
\label{tab:crystaldata}
\begin{tabular}{ l l } \hline\hline
Composition & UAsS \\ 
Space group & $P$4$_2$/$nmm$  \\ 
Pearson symbol & $tP$6 \\ 
Formula units per cell, Z & 2\\ 
Lattice parameters\footnote{The lattice parameter values were extracted from the powder x-ray diffraction data, see Table \ref{tab:latpar} (batch NZs 255).} & \\ 
$a$ / \AA & 3.8774(1)  \\ 
$c$ / \AA & 8.1618(4)  \\ 
$V$ / \AA$^3$ & 122.71(1) \\ 
Calc. density / g cm$^{-1}$ & 9.338\\ 
Range in \textit{h,k,l} & --7 $\leq$ $h$ $\leq$ 7 \\ 
&  --4 $\leq$ $k$ $\leq$ 7 \\ 
& --15 $\leq$ $l$ $\leq$ 8 \\ 
Absorption coeff. / mm$^{-1}$ & 79.950\\ 
T(max)/T(min) & 2.43\\ 
N($hkl$) measured & 2402\\ 
N($hkl$) unique & 307\\ 
R$_{int}$ & 0.0484\\ 
N($hkl$) observed & 297\\ 
Observation criterion & {\textit{F(hkl)} $\geq$ 4$\sigma$ [\textit{F(hkl)}]} \\
Refined parameters & 10\\ 
$R_1$ & 0.0282\\ 
$wR_2$ & 0.0585\\ 
Residual peaks / e \AA$^{-3}$ & --3.85 / 3.64\\ \hline\hline
\end{tabular}
\end{footnotesize}
\end{table}

\begin{table}[!ht]
\begin{footnotesize}
\caption{Atomic coordinates and equivalent displacement parameters (in Å$^2$) in UAsS.}
\label{tab:coordinates}
\begin{tabular}{c c l l l l} \hline\hline
Atom & Wyckoff site & $x/a$ & $y/b$ & $z/c$ & $U$$_{eq}$ \\ \hline
U & 2\textit{c} & \nicefrac{1}{4} & \nicefrac{1}{4} & 0.71508(5) & 0.00706(14)\\ 
As & 2\textit{a} & \nicefrac{3}{4} & \nicefrac{1}{4} & 0 & 0.00721(19)\\ 
S & 2\textit{c} & \nicefrac{1}{4} & \nicefrac{1}{4} & 0.3664(3) & 0.0059(4)\\ \hline\hline
\end{tabular}
\end{footnotesize}
\end{table}

\begin{table}[!ht]
\caption{Interatomic distances in the UAsS structure.}
\label{tab:distances}
\begin{tabular}{c l r  c l r  c l r} \hline\hline
\multicolumn{2}{c}{Atoms} & $\delta$, \AA & \multicolumn{2}{c}{Atoms} & $\delta$, \AA & \multicolumn{2}{c}{Atoms} & $\delta$, \AA\\ \hline
{U--}  & 4S & 2.821(6) &   {As--} & 4As & 2.742(1) &  {S--} & 4U & 2.821(6) \\
       & 1S & 2.846(3) &          & 4U  & 3.027(3) &        & 1U & 2.846(3) \\
       & 4As & 3.027(3) \\
  \hline\hline
\end{tabular}
\end{table}

\begin{table}[!ht]
\begin{footnotesize}
\caption{Lattice parameters and preparation method of UAsS in this and previous works\cite{Hulliger1968, Zygmunt1972, Zygmunt1974, Henkie1998, Henkie2001}.}
\label{tab:latpar}
\begin{tabular}{l c c c c c} \hline \hline
Batch number & $a$, \AA & $c$, \AA & $V$, \AA$^3$ & $c/a$ & Preparation method \\ \hline \hline

\multirow{2}{*}{UAsS (NZs181)}\footnote{LaB$_6$ was used as an internal standard for the refinement.} & \multirow{2}{*}{3.8788(1)} & \multirow{2}{*}{8.1649(5)} & \multirow{2}{*}{122.84(3)} & \multirow{2}{*}{2.105} & CVT in a vertical furnace, 950$^\circ$C, \\ 
& & & & & AsI$_3$ as a transport agent \\ \hline

\multirow{2}{*}{UAsS (NZs193)}$^a$ & \multirow{2}{*}{3.8761(1)} & \multirow{2}{*}{8.1681(4)} & \multirow{2}{*}{122.72(0)} & \multirow{2}{*}{2.107} & CVT in a vertical furnace, 950$^\circ$C, \\ 
& & & & & AsI$_3$ as a transport agent \\ \hline

\multirow{2}{*}{UAsS (NZs196)}$^a$ & \multirow{2}{*}{3.8756(1)} & \multirow{2}{*}{8.1627(3)} & \multirow{2}{*}{122.60(7)} & \multirow{2}{*}{2.106} & CVT, 880$^\circ$C$\rightarrow$950$^\circ$C, \\ 
& & & & & AsI$_3$ as a transport agent \\ \hline

\multirow{2}{*}{UAsS (NZs238)}$^a$ & \multirow{2}{*}{3.8763(1)} & \multirow{2}{*}{8.1620(3)} & \multirow{2}{*}{122.64(1)} & \multirow{2}{*}{2.106} & CVT, 900$^\circ$C$\rightarrow$970$^\circ$C, \\
& & & & & AsI$_3$ as a transport agent \\ \hline

\multirow{2}{*}{UAsS (NZs255)}$^a$ & \multirow{2}{*}{3.8774(1)} & \multirow{2}{*}{8.1618(4)} & \multirow{2}{*}{122.70(8)} & \multirow{2}{*}{2.105} & CVT, 890$^\circ$C$\rightarrow$980$^\circ$C, \\
& & & & & AsI$_3$ as a transport agent \\ \hline

\multirow{2}{*}{UAsS\cite{Hulliger1968}} & \multirow{2}{*}{3.874(3)} & \multirow{2}{*}{8.158(5)} & \multirow{2}{*}{122.46(1)} & \multirow{2}{*}{2.106} & CVT, 900-950$^\circ$C$\rightarrow$1000$^\circ$C, \\ 
& & & & & Br or I as a transport agent \\ \hline

\multirow{2}{*}{UAsS\cite{Zygmunt1972, Zygmunt1974}} & \multirow{2}{*}{3.884(5)} & \multirow{2}{*}{8.176(6)} & \multirow{2}{*}{123.37(9)} & \multirow{2}{*}{2.105} & CVT, 900-950$^\circ$C$\rightarrow$1000$^\circ$C, \\ 
& & & & & Br or I as a transport agent \\ \hline

\multirow{2}{*}{UAsS\cite{Henkie1998}} & \multirow{2}{*}{3.860(1)} & \multirow{2}{*}{8.128(2)} & \multirow{2}{*}{121.11(3)} & \multirow{2}{*}{2.106} & CVT, 880-920$^\circ$C$\rightarrow$950$^\circ$C, \\ 
& & & & & Br as a transport agent \\ \hline

\multirow{2}{*}{UAsS\cite{Henkie2001}} & \multirow{2}{*}{3.860(1)} & \multirow{2}{*}{8.128(1)} & \multirow{2}{*}{121.11(2)} & \multirow{2}{*}{2.106} & CVT, 880-920$^\circ$C$\rightarrow$950$^\circ$C, \\ 
& & & & & Br as a transport agent \\ \hline \hline

\end{tabular}
\end{footnotesize}
\end{table}

The single crystals of UAsS were analyzed by energy-dispersive X-ray spectroscopy with a Jeol JSM 6610 scanning electron microscope equipped with an UltraDry EDS detector (Thermo Fisher NSS7). The semi-quantitative analysis was performed with a 30 keV acceleration voltage. No impurity elements were detected, confirming that no reaction with the container took place during synthesis. The experimentally determined element ratio of 31.1 at.\% U: 35.2 at.\% As: 33.7 at.\% S was in good agreement with the equiatomic composition (33.3\% : 33.3\% : 33.3\%), obtained from the structure refinement. No evidence of twinning was found from the metallographic analysis or single crystal diffraction.

Calculations of chemical bonding were performed with the all-electron, local orbital full‒potential method (FPLO\cite{y23} (local density approximation as Perdew--Wang parametrization\cite{y24}) spin-polarized scalar relativistic calculation, standard basis set, 12×12×12 $k$ points). The lattice parameter and atomic coordinates were obtained from the experimental crystal structure refinement.

\begin{figure*}
    \centering
    \includegraphics[width=\textwidth]{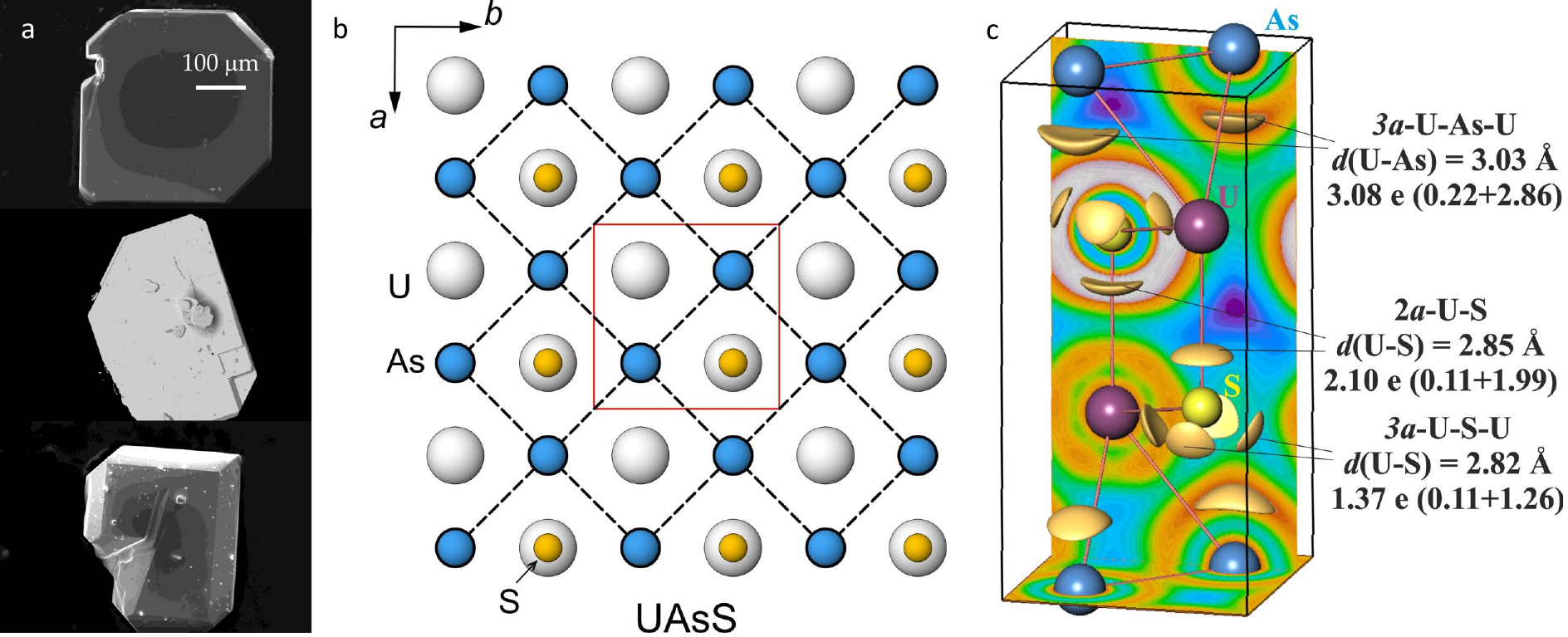}
    \caption{(a) Single crystals of UAsS typically have plate-like morphology with the large plane corresponding to the $ab$ plane. (b) Crystal structure of UAsS within the $ab$ plane. The As network differs from the zigzag or trapezoid chains observed in SmAsS and LaAsS, described previously\cite{Robredo2025}. (c) The bonding analysis of UAsS indicates that the bonds are the weakest between two U--S layers and between the U--S and As layers, suggesting a likely cleaving plane.}
    \label{fig:LP}
\end{figure*}

For the analysis of chemical bonding in position space, the electron localizability approach was utilized\cite{y38}. The electron density (ED) and the electron localizability indicator (ELI-D) were calculated with a specialized module implemented in the FPLO program package\cite{y25}. The topological features of the computed distributions of ED and ELI-D were analyzed with the program DGrid\cite{y39}. Further information about the bonding between atoms is obtained from the electron-localizability approach, combining the analysis of ED and ELI-D\cite{y38}. Analysis of ELID reveals three main atomic interactions in UAsS -- see Fig.~\ref{fig:LP}c: 

\begin{enumerate}
    \renewcommand{\labelenumi}{\textit{\roman{enumi}}.}
\item the 3-atomic U--S--U strongly polar interaction (only 8\% of the bond basin population originates from both U atoms) in the layer perpendicular to the $c$ axis;
\item the two-atomic U--S along the $c$ axis with the main contribution of S to the bond basin population (95\%), i.e., it is rather a "lone pair" on sulfur;  
\item the 3-atomic U--As--U, with the As main contributions to the bond basin population (93\%), i.e., it is also more lone-pair-like on As.
\end{enumerate}

Assuming that the material usually breaks along the planes formed by lone-pairs, the planes perpendicular to the $c$ axis are the most probable cleavage planes. The first one is between the U--S and As layers, separated by the "lone-pairs" on As is terminated by As atoms on one side (see Fig.~\ref{Figure S1:STM}c) and the U atoms on the other one (see Fig.~\ref{Figure S1:STM}d). The other one is between the U--S layers separated by the "lone-pairs" on S (interaction $ii$) is terminated by S atoms -- see Fig.~\ref{Figure S1:STM}d. For the visualization of the bonding analysis see Fig.~\ref{fig:LP}c.

\vspace{-10pt}
\section{Scanning transmission electron microscopy}\label{STEM}

\begin{figure*}
    \centering
    \includegraphics[width=\linewidth]{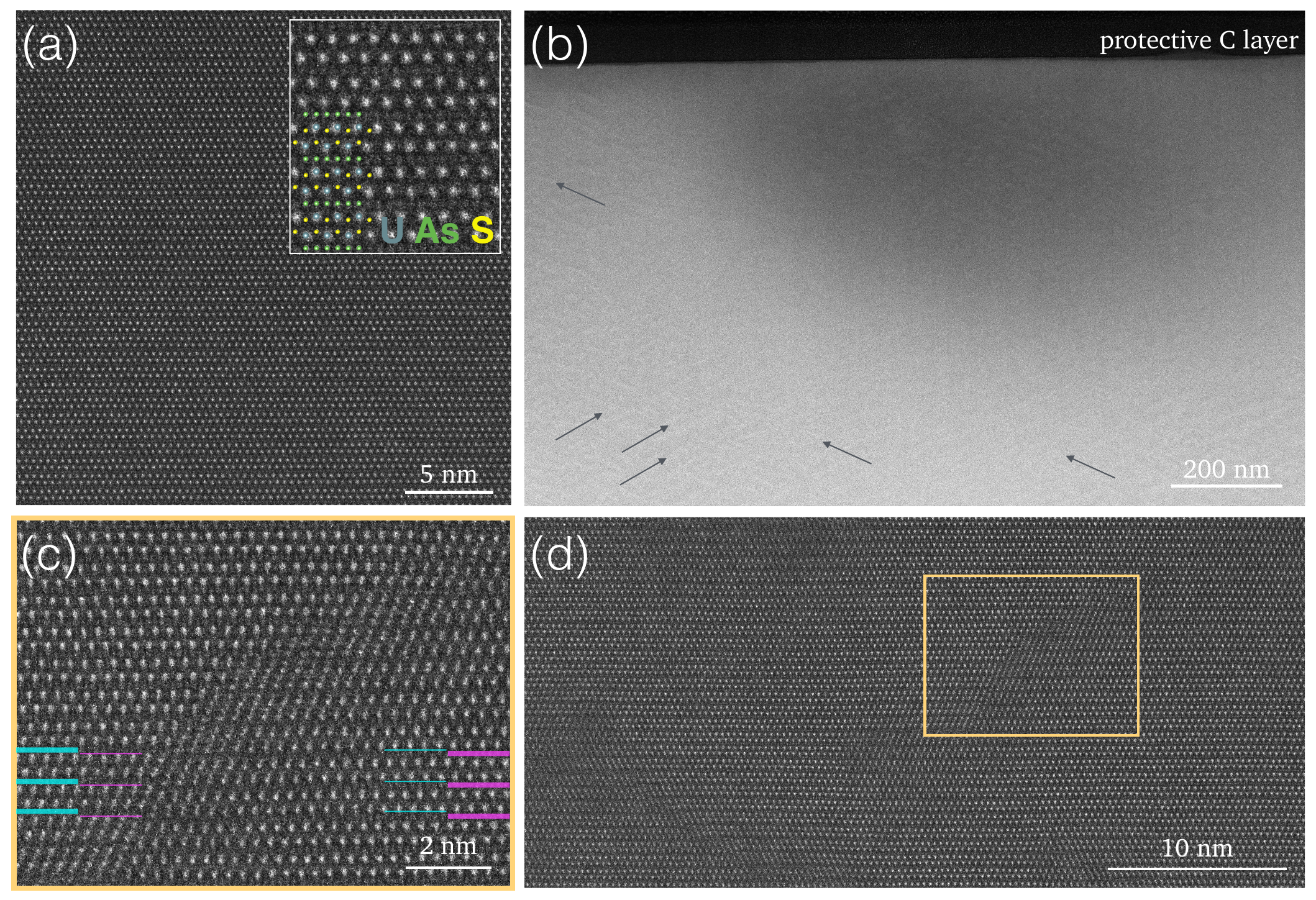}
    \caption{STEM characterization of UAsS single crystal. (a) Atomic-resolution HAADF-STEM image showing the [100] zone-axis projected crystal structure with (inset) atomic model overlay. Heavy U (grey-blue) and As (green) atomic columns are clearly resolved in the high-angle collection geometry, while light S (yellow) columns have significantly lower contrast. (b) Low-angle annular dark-field STEM imaging across a large field of view reveals global distribution of bright ``hatch-marks", indicating regions of high local crystalline strain which result in increased diffraction contrast. The slowly varying background intensity arises from thickness differences in the thinned FIB lamella. (c) and (d) High-resolution investigation of one ``hatch-mark" reveals crystalline stacking faults: small offsets in the $ab$ atomic planes along $c$. The cyan (magenta) lines in (c) highlight the vertical position of US planes on the left (right) side of a fault. Thick lines mark the As planes, the thin lines mark the same position extended to the other side of the fault.}
    \label{fig:stem}
\end{figure*}

A cross-sectional lamella for scanning transmission electron microscopy (STEM) analysis was prepared using the standard focused ion beam (FIB) lift-out procedure using a Thermo Fisher Scientific Helios NanoLab G5 FIB.

STEM images were collected on a double aberration-corrected JEOL ARM300F operating at 300 kV with a probe convergence semiangle of 30 mrad, in high-angle annular dark-field (HAADF, collection angles 68 to 280 mrad) and low-angle annular dark-field (LAADF, collection angles 27 to 110 mrad).

\section{Bulk physical properties} \label{bulk}

\begin{figure*}
    \centering
    \includegraphics[width=\linewidth]{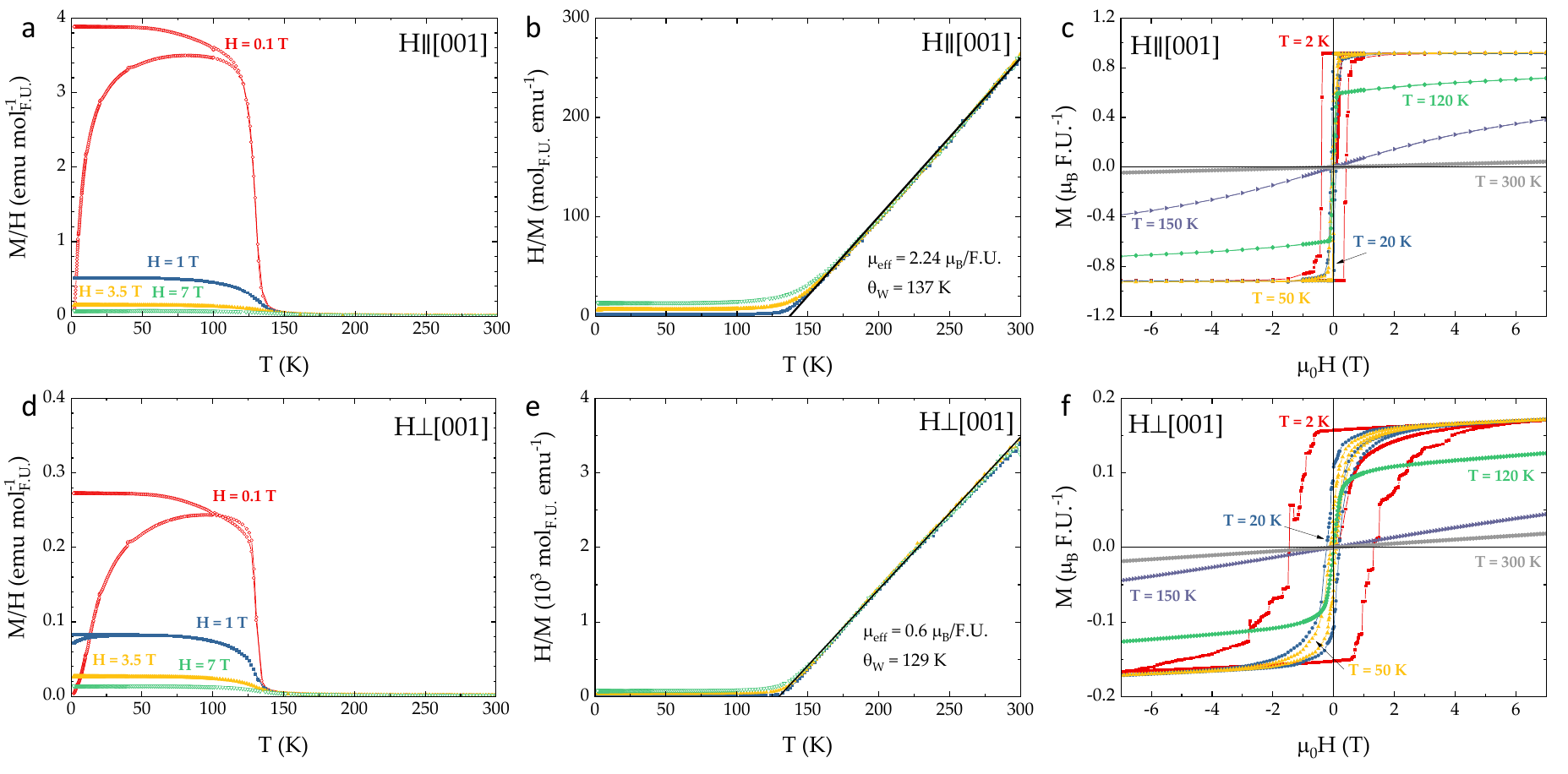}
    \caption{Magnetic properties of bulk UAsS as a function of temperature $T$ and field $H$ for $H \parallel [001]$ (a)-(c) and $H \perp [001]$ (d)-(f). A clear anisotropy is evident, as the size of both saturated and paramagnetic magnetic moments are significantly larger along the [001] easy axis.}
    \label{fig:BulkFM}
\end{figure*}

Magnetic properties were examined using a Quantum Design (QD) Magnetic Property Measurement System in the temperature range of 1.8–300 K and under various applied magnetic fields. Anisotropic measurements of UAsS single crystals are summarized in Fig.~\ref{fig:BulkFM}. Given crystal morphology, measurements have been carried out with the field perpendicular and parallel to the plane of the platelets, $i.e.$ the $c$-axis. 

AC electrical resistivity measurements were performed on a  QD Physical Property Measurement System, using a standard four-probe technique at temperatures between $T = 2$ and 300 K in $H = 0$  magnetic field. A current pulse of 0.01 mA with frequency 93 Hz for 1 s was applied along the $ab$-plane. 

\begin{figure*}
    \centering
    \includegraphics[width=0.7\linewidth]{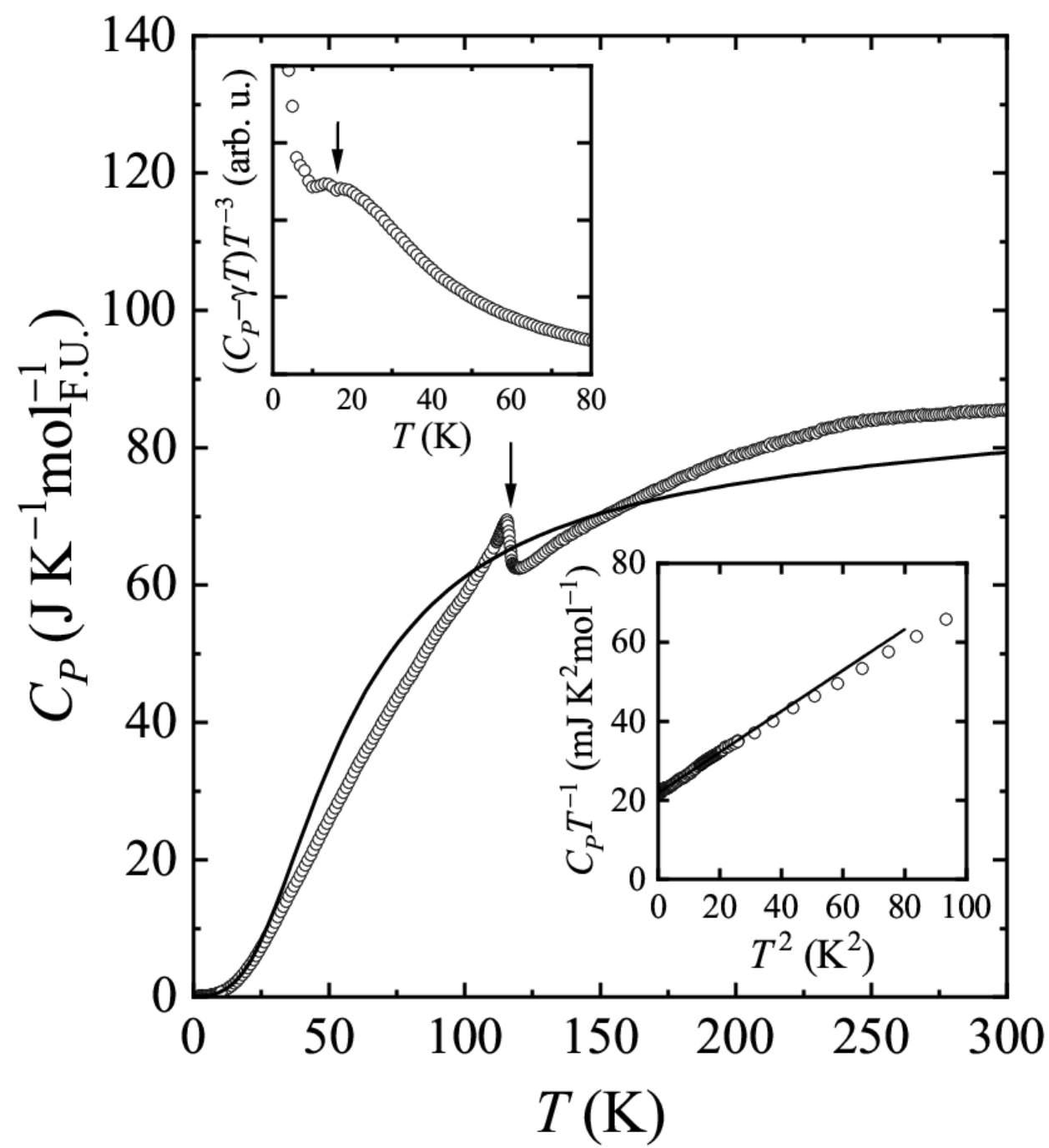}
    \caption{Specific heat $C_P$ of UAsS as a function of temperature $T$; the solid line represents a simulated curve based on the full Debye model using the parameters $\gamma$ and $\Theta_{\mathrm{D}}$, obtained from the low-temperature fit to Eq. (\ref{T3-Debye}). The arrow indicates the magnetic ordering temperature. Lower inset: low-temperature plot of $C_P/T$ versus $T^2$; the solid line is a fit of Eq. (\ref{T3-Debye}) to the experimental data. Upper inset: $C_P - \gamma T$ plotted as a function of temperature in the form $(C_P - \gamma T)/T^3$; the arrow highlights a possible bosonic anomaly, potentially originating from Einstein-like lattice vibrations.}
    \label{fig:Cp}
\end{figure*}

The specific heat data were collected on a QD PPMS from 0.4 K to 10 K and under various applied magnetic fields. The temperature dependence of the specific heat $C_P(T)$ of UAsS (see Fig.~\ref{fig:Cp}) is characteristic of intermetallic systems, with a dominant phononic background corresponding to a Debye temperature on the order of room temperature. A pronounced $\Lambda$-shaped anomaly at $\sim$115–120 K indicates a magnetic phase transition, with the temperature being consistent with the other physical properties.

Due to the lack of a nonmagnetic analogue and the absence of phonon density of states data from neutron scattering, the $f$-electron contribution to $C_P(T)$ could not be reliably extracted over the full temperature range. Analysis was therefore limited to the low-temperature regime ($T \lesssim 5$ K), where $C_P(T)$ is expected to be dominated by conduction electrons and acoustic phonons. In this regime, the data are well described by\cite{Gopal2012}:

\begin{equation}
\label{T3-Debye}
C_P(T) = \gamma T + \beta T^3,
\end{equation}

\noindent with $\gamma$ and $\beta$ representing the electronic and phononic coefficients, respectively. A linear fit to $C_P/T$ vs. $T^2$ (see Fig.~\ref{fig:Cp}, lower inset) yields $\gamma = 0.022(1)\,\mathrm{J\,K^{-2}\,mol^{-1}}$ and $\Theta_D = 224(1)$ K, using the standard relation:

\begin{equation}
\Theta_D = \left( \frac{12 \pi^4 r R}{5 \beta} \right)^{1/3},
\end{equation}

\noindent where $r$ is the number of atoms per formula unit and $R$ is the molar gas constant. However, these parameters are not applicable at higher temperatures, as the phonon contribution calculated from the full Debye model (see e.g. Ref.\cite{Gopal2012}) using the obtained $\Theta_{\rm D}$ already overestimates the experimental $C_P(T)$ data below approximately 150 K (see the solid line in Fig.~\ref{fig:Cp}). This discrepancy indicates that the fitted $\beta$ coefficient likely includes additional low-energy contributions beyond the phononic term, implying that the intrinsic Debye temperature of the lattice is in fact higher. Consequently, this simplified description based solely on lattice and electronic terms becomes increasingly inadequate at elevated temperatures, where other effects -- such as magnetic excitations (e.g., magnons) or many-body interactions (e.g., the Kondo effect) -- may play a non-negligible role. Ultimately, the lack of a reliable reference for the phononic background brings us back to the central limitation of this analysis: the inability to disentangle the individual contributions to the specific heat beyond the low-temperature regime.

To explore the possible presence of localized Einstein-type vibrational modes, the quantity $(C_P(T) - \gamma T)/T^3$ was plotted as a function of temperature (see Fig.~\ref{fig:Cp}, upper inset). In this representation, the Debye contribution appears as a constant background, while any additional Einstein-like bosonic excitation is expected to manifest as a broad maximum centered around $T \approx \Theta_E/5$. This behavior can be understood from the temperature dependence of the specific heat in the Einstein model\cite{Gopal2012}:

\begin{equation}
    C_E(T) = 3rR \left( \frac{\Theta_E}{T} \right)^2 \frac{e^{\Theta_E/T}}{\left(e^{\Theta_E/T} - 1\right)^2},
\end{equation}

\noindent where $\Theta_E$ is the Einstein temperature, $R$ is the molar gas constant, and $r$ is the number of atoms participating in this type of vibrational mode. As seen in Fig.~\ref{fig:Cp}, a weak maximum is observed near 16 K, which could tentatively be attributed to a vibrational mode with $ \Theta_E \approx 80\,\mathrm{K}$. However, due to the limited reliability of the phononic background estimate, a quantitative analysis of this feature is not feasible.

\section{Magneto optical Kerr} \label{Kerr}

\begin{figure*}
    \centering
    \includegraphics[width=\linewidth]{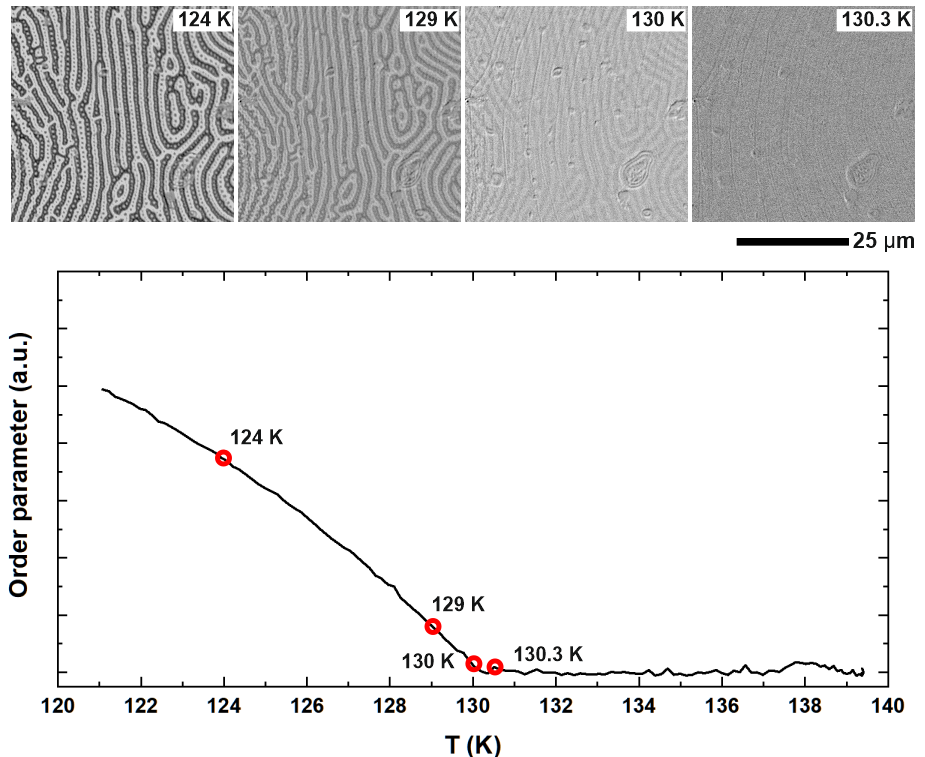}
    \caption{To determine the transition temperature, we introduced an ordering parameter based on the standard deviation (STD) of the image pixel intensity from the average gray level, which illustrates the appearance of the domain contrast at 130 K. The correspondent Kerr images, obtained above and below illustrate the transition.}
    \label{fig:MD3}
\end{figure*}

The magnetic domains, observed by polar Kerr microscopy on the basal plane of the UAsS single crystal in the ferromagnetic state, are shown in Fig.~\ref{fig:BulkFM}. At a temperature of 120 K, i.e. just below the transition temperature of about 130 K (see Fig.~\ref{fig:MD3}), the sample exhibits a two-phase branched domain structure that is typical of magnetic materials with a quality factor (ratio of anisotropy energy to stray field energy) much larger than 1, i.e. a strong and dominating uniaxial anisotropy perpendicular to the plane of observation. The domains at 120 K actually resemble the first-generation branched domains found on the basal plane of a NdFeB crystal with a thickness around 40 µm (compare Fig. 5.5.b in ref. \cite{MD1}). If the quality factor of our UAsS material would be of the same order as that of NdFeB (about 4), a multiple generation branched pattern would be expected given a crystal thickness of 0.1 mm in our case. The presence of just a first-generation pattern indicates that for our material the quality factor and thus uniaxial anisotropy must be higher than that of NdFeB material.

\begin{figure*}
    \centering
    \includegraphics[width=\linewidth]{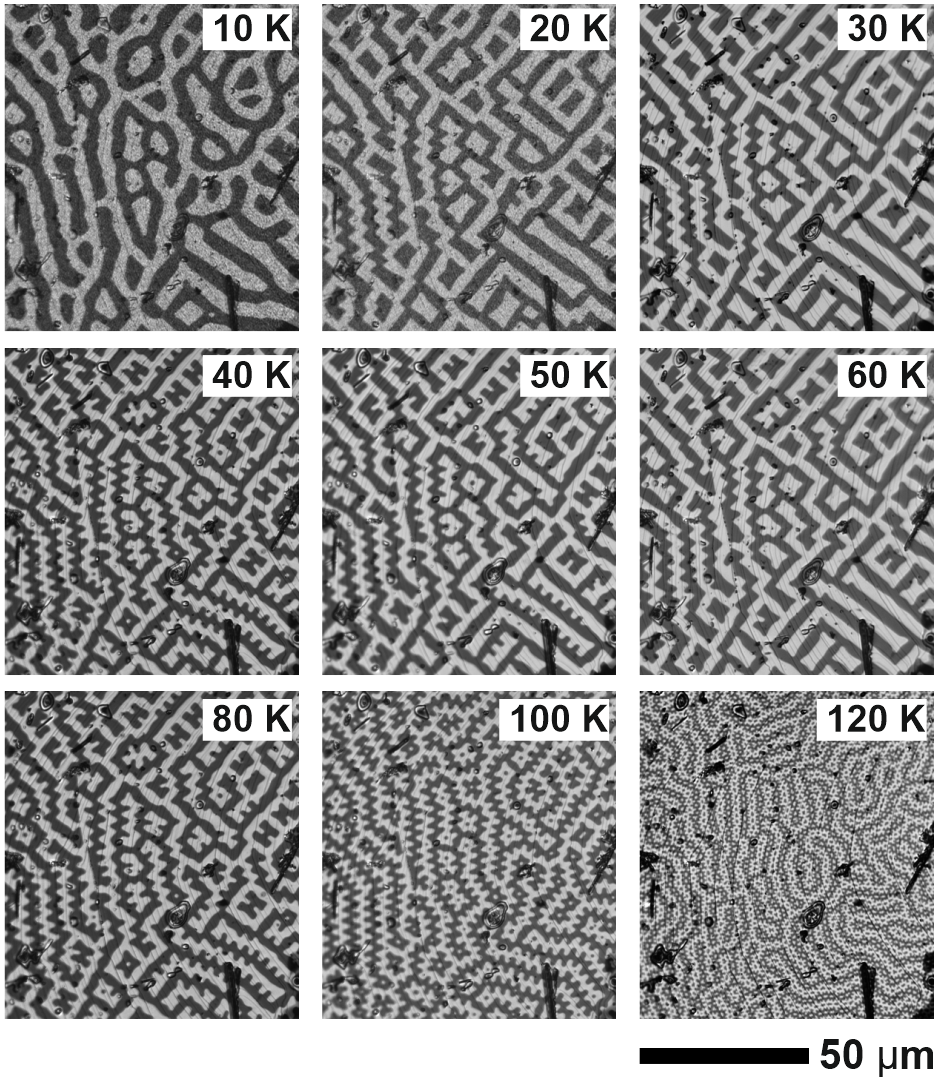}
    \caption{Magnetic domains observed on the basal plane of the UAsS single crystal in the ferromagnetic state at different temperatures demonstrate “easy axis” and “easy cone” type of magnetic anisotropies at high temperatures and at low temperatures, respectively, indicating the presence of a spin reorientation transition.}
    \label{fig:MD2}
\end{figure*}

The evolution of the domain structure at decreasing temperature -- characterized by the suppression of branching, the formation of straight domain walls, and an increase in domain width -- may be associated with a phase transition from a high-temperature “easy axis” phase to a low-temperature “easy cone” type of magnetic anisotropy phase (spin reorientation). This behavior is similar to that reported in Nd$_2$Fe$_{14}$B\cite{MD2, MD3} or rare-earth based magnets\cite{MD4, MD5}, where it is attributed to the emergence of an additional planar anisotropy coefficient. The evolution of magnetic domains at finer temperature steps can be found in the Fig.\ref{fig:MD2}.

\section{Point contact spectroscopy} \label{PCS}

Point-contact (PC) spectroscopy is a powerful method for studying quasiparticle excitations, which can interact with conducting electrons in solids \cite{Naidyuk2005}. Depending on the regime of current flow through the PC -– ballistic, diffusive, thermal –- different phenomena can be studied using PC spectroscopy: electron-phonon interaction, Kondo effect, crystal-electric field excitations, or different phase transitions. It is important to note that extremely high current density and electric field can be realized in PCs and a strong nonequilibrium state of electrons can be reached, the relaxation of which leads to the spectroscopic information, or leads to thermal heating. In the last case, a resistivity-dependent phenomenon is clearly observed in derivatives of current-voltage characteristics of PCs.

\begin{figure*}
    \centering
    \includegraphics[width=\linewidth]{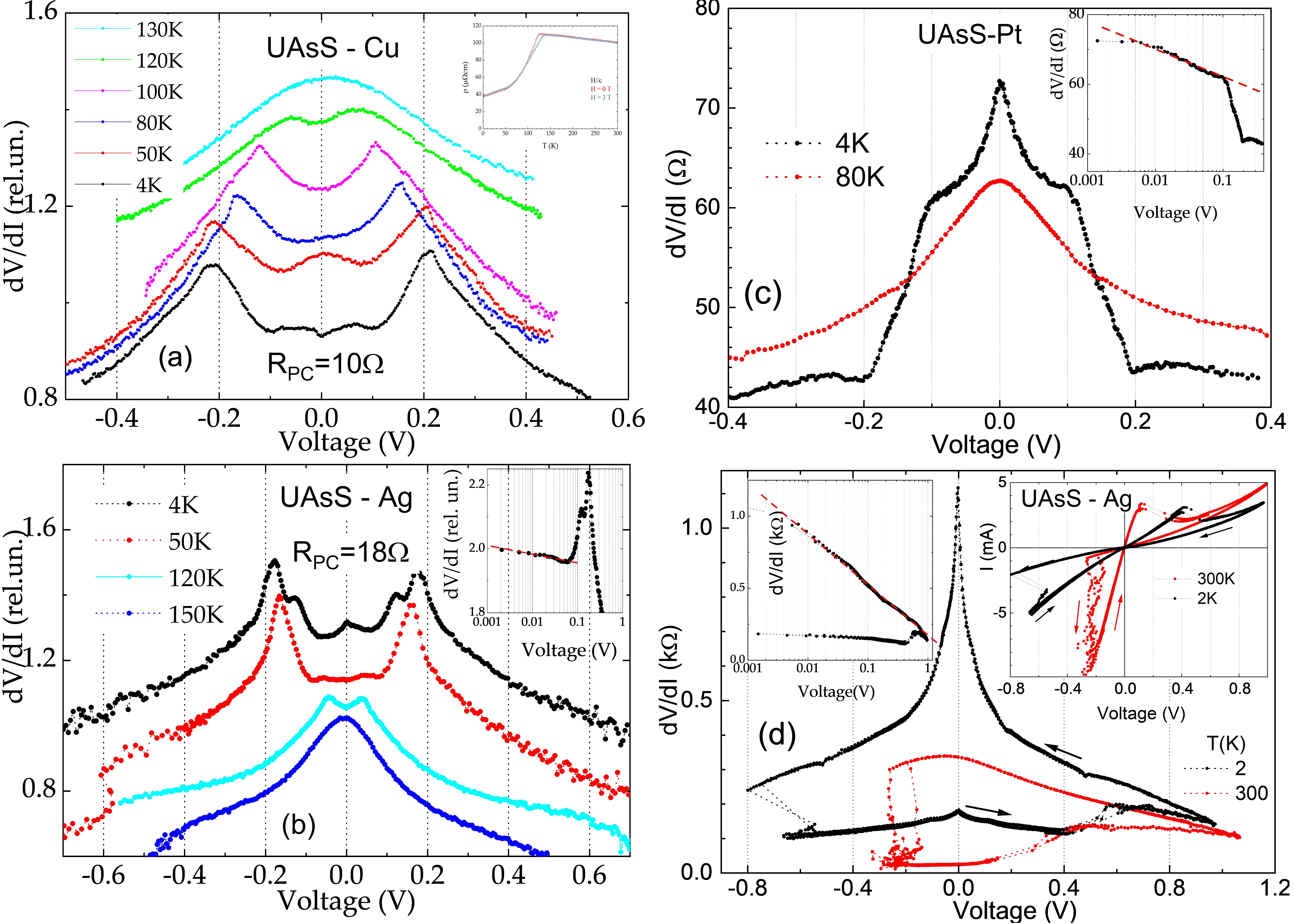}
    \caption{(a) $dV/dI(V)$ of UAsS--Cu PC at different temperatures. All curves at temperatures above 4 K are shifted downwards for clarity. The inset shows specific resistivity data. (b) $dV/dI(V)$ for UAsS–Ag PC at different temperatures. All curves at temperatures above 4 K are shifted upwards for clarity. The inset shows $dV/dI(V)$ at 4 K on the log scale. (c) $dV/dI(V)$ with a dominating Kondo-like zero-bias peak of UAsS–Pt PC at different temperatures. The inset shows $dV/dI(V)$ at 4 K on the log scale. (d) $dV/dI(V)$ of UAsS--Ag PC with the switching effect at helium and room temperatures. The right inset shows the $I(V)$ curves for PC from the main panel. The left inset shows $dV/dI(V)$ at 2 K on the log scale.}
    \label{fig:PS1}
\end{figure*}

Point contacts on the surface of single-crystal UAsS were established with a thin wire of elemental metals (Ag, Pt, Fe, Co) that touched the $ab$-plane of the sample. We measured the current–voltage $I(V)$ and the differential resistance ($dV/dI(V)$) characteristics of PCs by sweeping the $dc$ current $I$ through PC, on which a small $ac$ current $i$ was superimposed, and detecting the first harmonic of ac signal using a standard lock-in technique in a four-probe configuration. To find the switching effect, we swept voltage back and forth across the point contact, increasing its amplitude until resistance switching was observed. The measurements were carried out in the temperature range from liquid helium up to room temperature and at a magnetic field up to 15 T.

Fig.~\ref{fig:PS1} shows $dV/dI(V)$ of PC on UAsS with two symmetrically located maxima around $V = 0$. The behavior of $dV/dI(V)$ (see, e.g., positive polarity) resembles the bulk resistivity $\rho(T)$ (see inset). The double maximum structure in $dV/dI(V)$ narrows with increasing temperature and disappears above 120 K, which demonstrates its connection with the magnetic transition. These observations testify in favor of the thermal regime in PC when PC temperature increases with a bias voltage \cite{Verkin1979}.

Another group of $dV/dI(V)$ spectra shows similar whole behavior (Fig.~\ref{fig:PS1}b) with an additional zero-bias maximum at the lowest temperature of 4 K. This maximum has a log-behavior, as can be seen from the inset in Fig.~\ref{fig:PS1}b, and could be due to the Kondo-scattering at the interface\cite{Naidyuk1982}. Overall behavior of dV/dI(V) above the magnetic transition temperature around 120 K is nonmetallic, which corresponds to vanishing $\rho(T)$ with a temperature rise. 

For PCs of higher resistance or lower size, it is expected that near-surface regions will have a major contribution to the PC conductivity. For such PCs, zero-bias or ”Kondo” maximum is more pronounced (Fig.~\ref{fig:PS1}c) and “magnetic” features are less visible as shoulders instead of peaks. Also, clear log-type dependence is seen in $dV/dI$ (see Fig.~\ref{fig:PS1}c, inset), which is characteristic of Kondo-type scattering in PCs \cite{Naidyuk1982}. Magnetic field up to 15 T had a negligible influence on such observed “magnetic” features.

Fig.~\ref{fig:PS1}d displays the resistive switching in PC. When the voltage reaches about + 0.4 V, $dV/dI(V)$ jumps above + 0.6 V and then, by further sweeping, transfers at about 1 V to the higher resistive state (HRS) with a sharp zero-bias peak. Then, at negative voltage around -0.8 V, $dV/dI(V)$ returns to the low resistive state (LRS). Two such repeatable cycles with switching between LRS and HRS are shown in Fig.~\ref{fig:PS1}d for two temperatures of 2 and 300 K. It can be seen that the switching is preserved up to room temperature. Note, that the pronounced sharp $dV/dI(V)$ peak for HRS has distinctly log-behavior for the lowest temperature of 2 K (left inset of Fig.~\ref{fig:PS1}d). The voltage range and shape of the switching loops are similar to that observed in PCs on a series of transition metal dichalcogenide crystals, such as MoTe$_2$, WTe$_2$, TaMeTe$_4$ (Me = Ru, Rh, Ir), TiSe$_2$, TiSe$_S$, Cu$_x$TiSe$_2$, VSe$_2$ and TiTe$_2$, which we investigated in Refs.\cite{Naidyuk2021, Bashlakov2023, Bashlakov2024, Kvitnitskaya2025}. Such similarity may indicate that the nature of the switching is connected with reversible modification/realignment of the crystal structure of the layered material in the PC core, apparently due to displacement/shift of ions under a high electric field.

Finally, we have conducted PC measurements on UAsS samples and have observed both magnetic transition and Kondo-type features in $dV/dI(V)$ characteristics. With increasing PC resistance, that is, with decreasing PC size, Kondo-type features have increased significantly while the magnetic transition peculiarities have weakened. Resistive switching was surprisingly observed in the studied PC both at helium and room temperatures, with a change in resistance by one order of magnitude similar. This indicates an increasing role of surface properties.

\section{Resistivity as a function of hydrostatic pressure} \label{Pressure}

\begin{figure}[b!]
\includegraphics[width=0.5\textwidth]{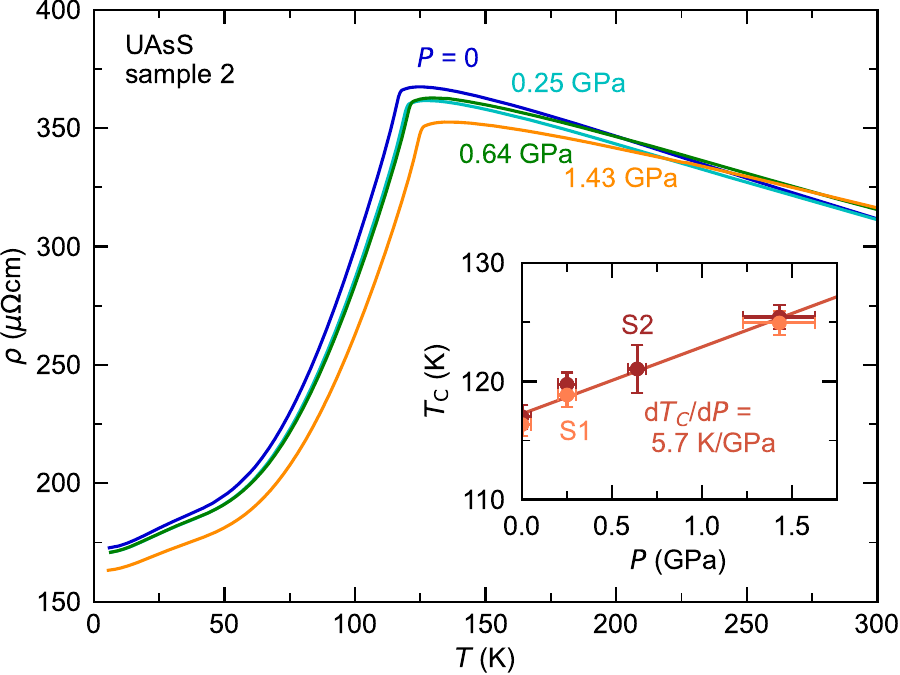}
\caption{Effect of hydrostatic pressure $P$ on the temperature dependence of resistivity $\rho$(T) of UAsS. The insert shows the pressure dependence of the ferromagnetic transition temperature $T_\mathrm{C}$.}
\label{fig:res_P}
\end{figure}

Fig.~\ref{fig:res_P} shows the temperature dependence of the resistivity $\rho (T)$ of UAsS under hydrostatic pressure from room temperature down to 6 K. At ambient pressure, $\rho (T)$ exhibits a kink at the ferromagnetic transition temperature $T_\mathrm{C} \approx $117 K. Under hydrostatic pressure, the transition linearly increases with a rate of 5.7 K/GPa. The overall temperature dependence of the resistivity persists up to 1.4 GPa. Hydrostatic pressure enhances the ferromagnetic order, and UAsS is not a candidate for a pressure-induced ferromagnetic superconductor as observed in other ferromagnetic systems -- for example, UGe$_2$ \cite{Saxena2000} and UAs$_2$ \cite{li2026}.

Two samples of UAsS were contacted with 25 $\mu$m gold wires using 4929 DuPont silver paste. We measured the resistance of the samples by the four-point method. The geometric factor was determined using a high-resolution image of the samples. We normalized the resistivity at room temperature to the mean value of both samples, $\rho_0 = (314 \pm 88)\,\mu\Omega$cm. The measurements were performed in a 5 mm CuBe/NiCrAl piston cylinder cell using polyethylsiloxane (PES-1) as pressure-transmitting media. The pressure in the cell was determined from the pressure dependence of the superconducting transition temperature of Sn, determined by the four-point method \cite{Eiling1981}. For the pressure point at 1.4 GPa, the pressure was determined with a different method, due to loss of contact to Sn. We used the linear relation of pressure and applied force, as well as pressure and cell length from the previous pressures. The measurements were performed in a QD Physical Property Measurement System (PPMS). We determined the resistance of the samples from 300 K to 6 K for warming and cooling with a sweep rate of 0.5 K/min (1 K/min for 0.64 GPa) to compensate for the thermal lag, which is caused by the large thermal mass of the pressure cell. The curves shown here have been corrected for this lag assuming that the actual sample temperature for a certain value of resistivity lies in the middle between the measured temperatures for this resistance value of the up and down sweeps.

\section{ARPES} \label{Arpes}

ARPES measurements on UAsS were performed on the (001) plane. The synchrotron experiments were carried out at BLOCH beamline in Max-IV Laboratory. Measurements were performed with angular resolution of 0.1$^o$. During the measurements the sample temperature was kept at 20 K, and the photon energy was varied within the range of 30-120 eV. The overall energy resolution was within 5 meV. Throughout the measurements the chamber pressure was kept at $7\times10^{-11}$ mBar.

\begin{figure}[ht!]
    \centering
    \includegraphics[width=0.8\linewidth,trim = 0 1cm 0 1cm]{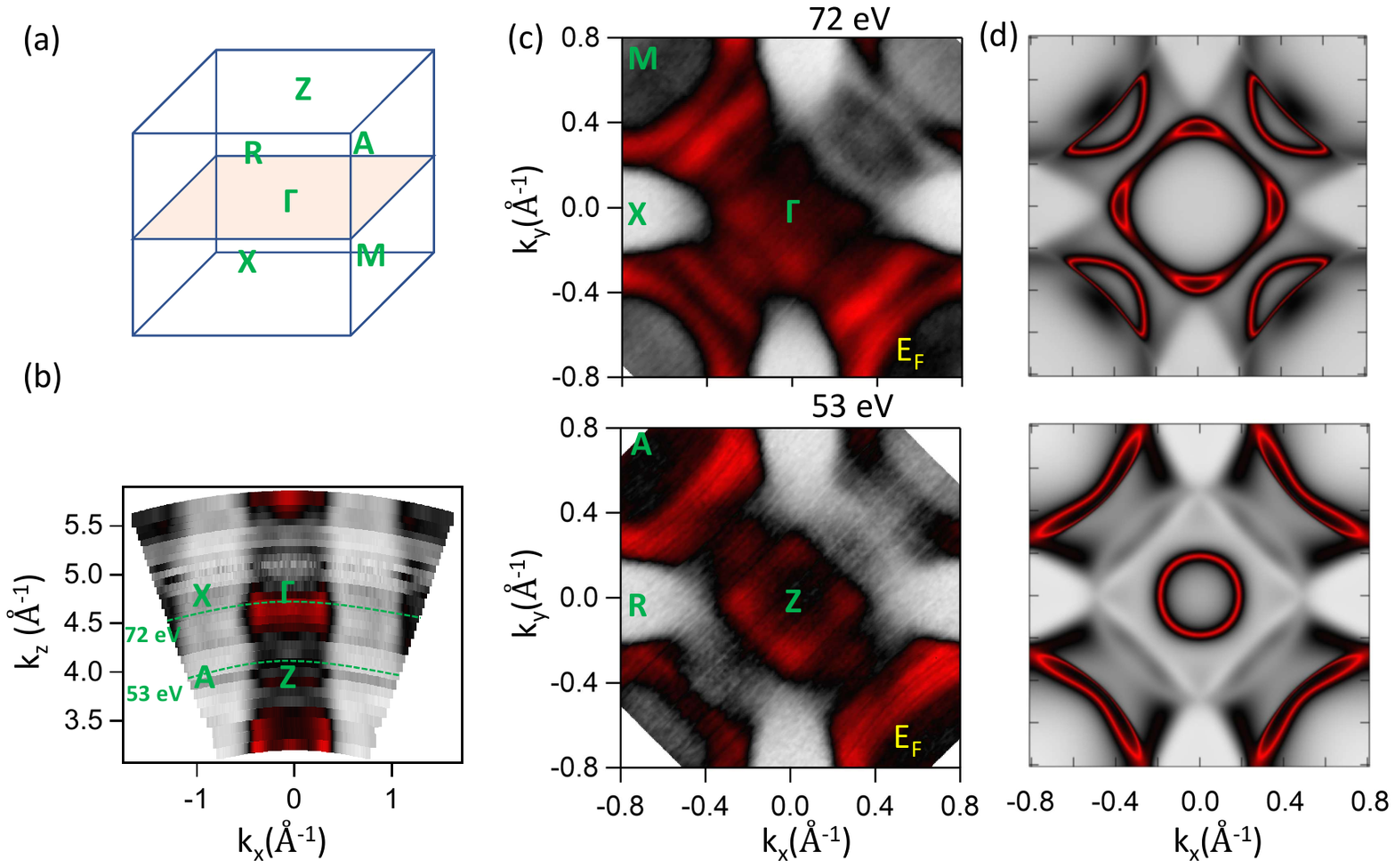}
    \caption{Identification of high-symmetry photon energies. (a) Three-dimensional BZ with high-symmetry points. (b) Out-of-plane FS map. (c) In-plane Fermi surface maps taken with $h\nu = 72$ eV and $53$ eV and (d) calculated FS maps at k$_z$=0 (top) and k$_z$=$\pi$ (bottom).}
    \label{Figure S1:ARPES}
\end{figure}


The Brillouin zone (BZ) with the high-symmetry points of UAsS is depicted in Fig.~\ref{Figure S1:ARPES}a. The out-of-plane FS map is presented in Fig.~\ref{Figure S1:ARPES}b. Following the overall symmetry of the ou-of-plane FS map, we identified the two high symmetry points at 72 eV and 53 eV. The in-plane FS maps recorded with $h\nu = 72$ eV and $53$ eV (Fig.~\ref{Figure S1:ARPES}c) are nicely reproduced by first principles DFT calculations at $k_z = 0$ and $k_z = \pi$, respectively (Fig.~\ref{Figure S1:ARPES}d). $72$ eV corresponds to 4.65 \AA$^{-1}$, which is $k_z\sim12\pi/c$ plane and $53$ eV to 4.1 \AA$^{-1}$ and $k_z \sim 11\pi/c$ plane.

\section{Details on the DFT calculations} \label{DFT}

\begin{figure*}
    \centering
    \includegraphics[width=0.8\linewidth]{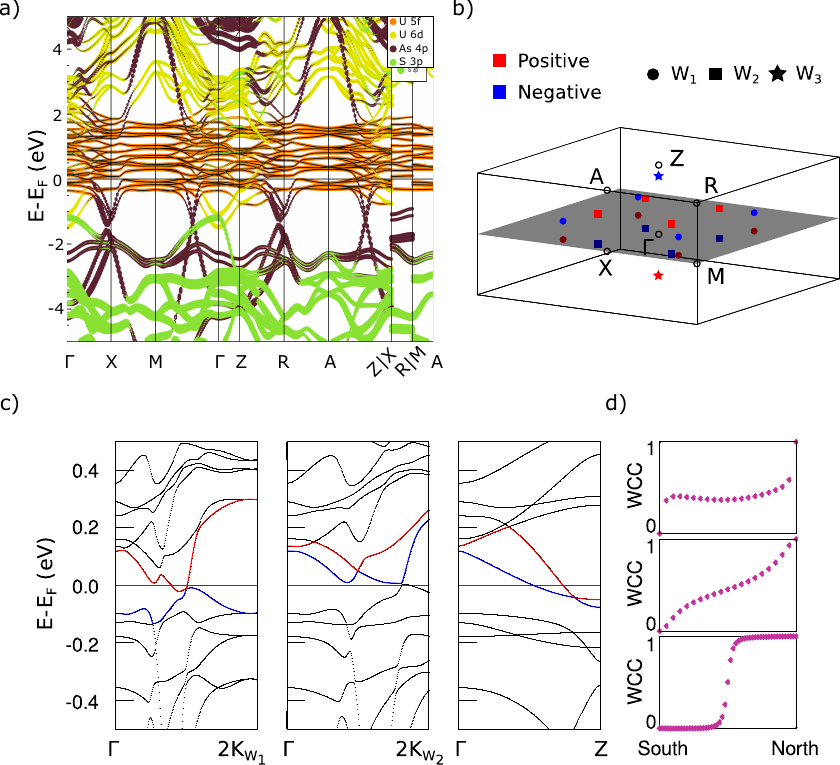}
    \caption{Bulk band structure and Weyl nodes close to the Fermi level. (a) Bulk band structure with colored orbital weights. In orange, we see the U $5f$ electrons occupying the Fermi level. Below the Fermi level, where the nodal lines lie, the weight comes mainly from As $4p$ orbitals, with U $5d$ and S $3p$ orbitals away from the Fermi level. (b) Position of Weyl nodes in the BZ. Red stands for positive topological charge and blue for negative. We denote with different symbols the Weyl nodes related by symmetry (same energy). (c) Band dispersion crossing the Weyl nodes. (d) Wannier charge center calculation on a sphere surrounding the Weyl nodes, showing all Weyls have charge 1.}
    \label{fig:fig_dft_2}
\end{figure*}

We performed DFT calculations as implemented in FPLO\cite{FPLO}. The local spin density approximation (LSDA) was employed for the exchange-correlation potential with the Perdew--Wang parameterization\cite{LSDA}, and the spin-orbit coupling (SOC) was considered based on the second variation method\cite{DFT-SOC}. A $\Gamma$-centered Monkhorst-Pack k-point grid of ($9 \times 9 \times 5$) was used for reciprocal space integration. We ensured convergence up to $10^{-5}$eV per unit cell. The magnetic groundstate is ferromagnetic, with $2.18\mu_B$ per U atom, consistent with our experimental fit. We constructed maximally localized Wannier functions from the DFT calculation as implemented in FPLO, and we run surface state calculations following the iterative Green's function method as implemented in WannierTools\cite{WannierTools}.

We identified three families of symmetry-related Weyl nodes, $W_1$, $W_2$ and $W_3$. They are located at positions $K_{W_1}=(0.428, 0.428, 0.058)$, $K_{W_2}=(0.270, 0.270, 0.094)$ and $K_{W_3}=(0, 0, 0.315)$ in \AA${}^{-1}$ units (see Fig.~\ref{fig:fig_dft_2}a), at $-11$ meV, $+51$ meV and $-42$ meV from the Fermi level. The first Weyl node is type-II, the second is type-I and the third showing a band dispersion in between type-I and type-II (see Fig.~\ref{fig:fig_dft_2}b). We computed the Wannier charge centers on a sphere surrounding each Weyl node. All of them have charge $\pm1$.

\section{Scanning tunneling microscopy/spectroscopy (STM/STS)} \label{STM}

\subsubsection{Experimental methods on STM/STS}\label{subsubsec:STMmethod}

Experiments were carried out in an ultra-high vacuum (UHV) system housing a commercially available STM (Unisoku USM1300) that is capable of operating at temperatures down to 0.36 K and high magnetic fields up to 11 T (orthogonal to the sample surface). All the STM/STS data shown in this work have been measured either at 4.2 K or 0.36 K, as indicated on each figure. Tunneling spectroscopy (d$I$/d$V$) data were acquired using standard lock-in amplifier techniques, where a modulation voltage ($V$$_{\mathrm{a.c.}}$) at a frequency of 833 Hz is applied to the sample bias voltage ($V$$_{\mathrm{s}}$) during data acquisition. Before each experimental run, STM-tip made of Pt/Ir was calibrated against the Shockley surface state of Cu(111). STM/STS data analyzing and post-rendering were performed using the freeware WSxM\cite{Horcas2007}.

Clean surfaces of UAsS were achieved by mechanically cleaving at low temperature (95 K) the samples in a differential background pressure of 1.6$\times$10$^{\mathrm{-10}}$ mbar. The cleavage is assisted by glueing beforehand a small stainless-steel nut on top of the sample surfaces in air using a UHV-compatible epoxy (EPO-TEK H20D).  After cleaving, the samples were directly transferred to the cold STM stage which is already precooled at 4.2 K. In this study, we successfully cleaved 6 samples, which yielded consistent STM/STS results as those shown in this work.

\subsubsection{Cleavage plane identification}\label{subsubsec:cleavage}

\begin{figure*}
    \centering
    \includegraphics[width=0.9\linewidth]{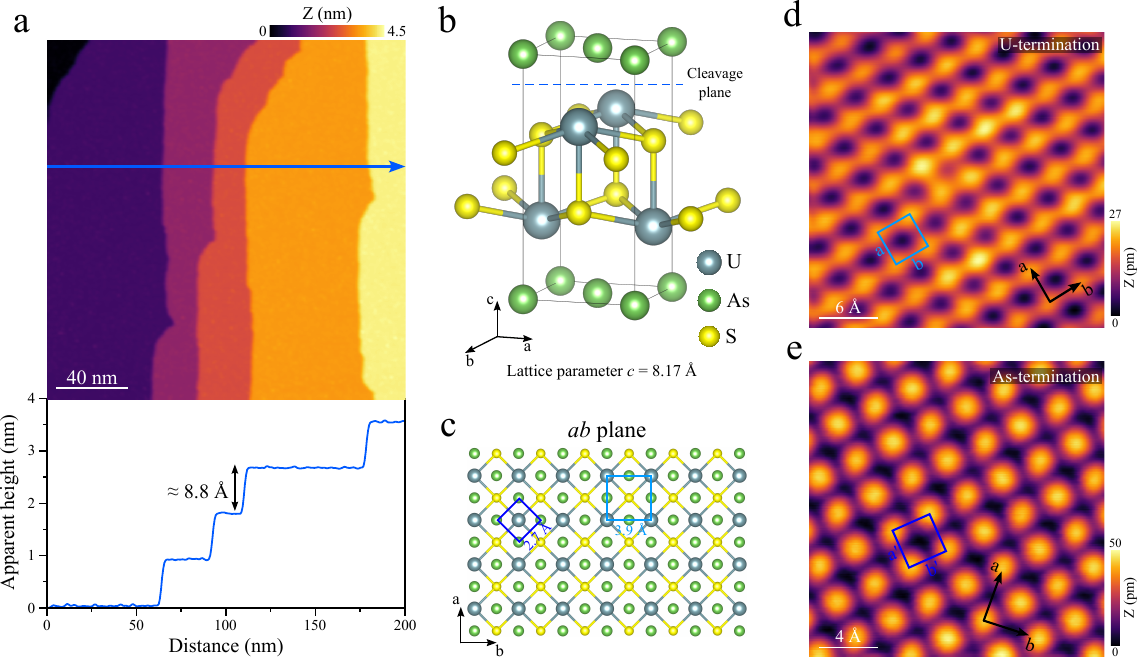}
    \caption{a) Large-scale STM topography images showing the as-cleaved surfaces of UAsS, where several atomic steps are visible and with a separation of $\approx$ 8.8~{\AA} between consecutive steps. b) Crystal structure of UAsS, where the main crystallographic axes are labelled. c) Top-view structural model showing the in-plane atomic arrangement of the ab-plane of UAsS. d) Atomically resolved STM topography image of the U-terminated surface, revealing a square atomic arrangement with a periodicity of $\approx$ 3.9~{\AA}. e) Representative atomically resolved STM topography image of the As-termination, which shows a square atomic arrangement with a periodicity of $\approx$ 2.7 ~{\AA}. Acquisition parameters: a) $V_{\mathrm{s}} = 1$ V, $I_{\mathrm{t}} = 20$ pA, $T$ = 4.2 K. d) $V_{\mathrm{s}} = 70$ mV, $I_{\mathrm{t}} = 0.2$ nA, $T$ = 4.2 K. e) $V_{\mathrm{s}} = 100$ mV, $I_{\mathrm{t}} = 50$ pA, $T$ = 4.2 K.}
    \label{Figure S1:STM}
\end{figure*}

To identify the cleavage plane, we recorded large-scale STM topography images (Fig.~\ref{Figure S1:STM}a) of the freshly cleaved samples of UAsS to know the separation between consecutive atomic planes. In this case, our results indicated an interplane distance of $\approx 8.8$ \AA, as can be deduced from the cross-sectional height profile (blue line). This experimental value is in good agreement with the unit cell parameter $c$ of UAsS (8.17 \AA, Fig.~\ref{Figure S1:STM}b), suggesting that the cleavage direction occurs parallel to the \textit{ab}-plane. Within this atomic plane, we may further distinguish between S-terminated, U-terminated, and As-terminated surfaces, with all of them displaying a square-like atomic symmetry but with varying in-plane periodicities (Fig.~\ref{Figure S1:STM}c). In our experiments, most of the measured surfaces show an in-plane periodicity of $\approx 3.88$ \AA (Fig.~\ref{Figure S1:STM}d). Based on the arguments discussed in the main text, these surfaces should correspond to an U-termination. We note here that the otherwise As-terminated surfaces have also been observed (Fig.~\ref{Figure S1:STM}d), although with a much lower probability, suggesting that it is not the most energetically favored cleavage plane. In the latter case, the surfaces shows a smaller in-plane periodicity of $\approx$ 2.7 \AA (Fig.~\ref{Figure S1:STM}e) in full agreement with results of crystal structure refinement (see Table \ref{tab:distances}).

\subsubsection{Electronic structure comparison between STS, ARPES and DFT}\label{subsubsec:comparison}

\begin{figure*}
    \centering
    \includegraphics[width=0.60\linewidth]{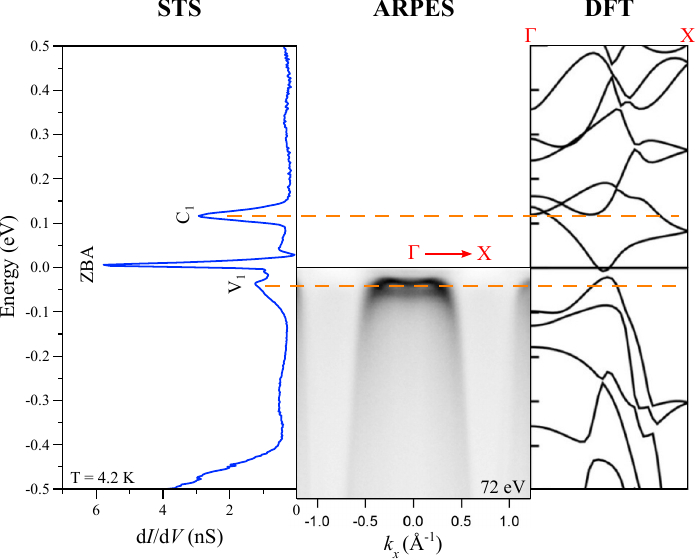}
    \caption{Comparison between the experimentally measured electronic structure from STS and the band structure probed by ARPES along $\Gamma$-X direction, as well as the ab-initio calculated band structure. STS acquisition parameter: $V_{\mathrm{a.c.}} = 3$ mV, $T$ = 4.2 K.}
    \label{Figure S2:STM}
\end{figure*}

In Fig.~\ref{Figure S2:STM} we compare the electronic structure recorded from local tunneling spectroscopy data with the band structure obtained from experimentally measured ARPES data, and also from DFT calculations. As can be seen, the ZBA and V$_1$ peak can be associated, respectively, with the non-dispersive flat band near EF and the first valence band observed in ARPES EDMs. However, in DFT calculation, there are slight differences. First, the ZBA can not be related to any band feature in the theoretical calculation, while the V$_1$ peak undergoes a severe band renormalization effect which pushes the associated bands closer to $E_{F}$ in the experimental data. For the unoccupied states, the C$_1$ peak can only be contrasted with DFT calculation, yielding consistent result.

\subsubsection{Absence of ZBA in the As-terminated surfaces}\label{subsubsec:absence}

\begin{figure*}
    \centering
    \includegraphics[width=0.8\linewidth]{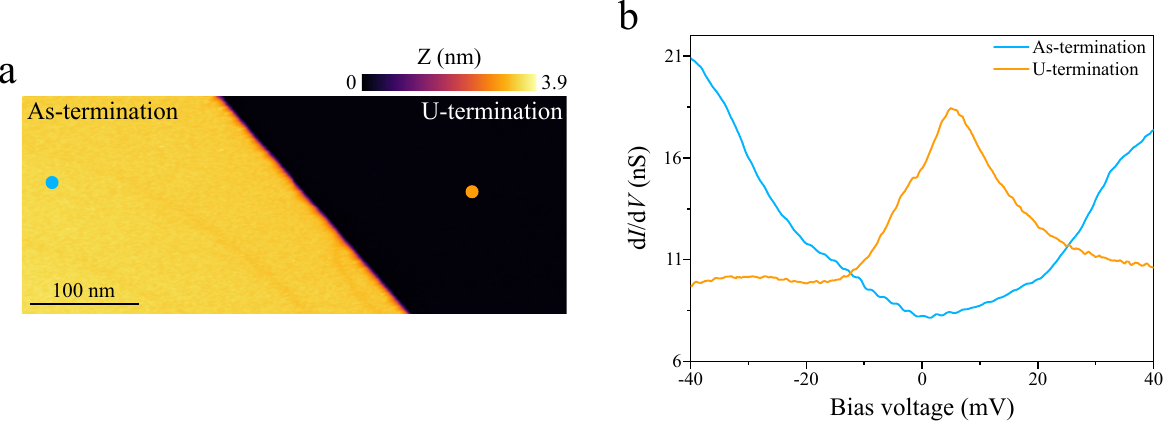}
    \caption{(a) Large-scale STM topography image showing two atomic terraces that correspond to the As-terminated and U-terminated surfaces. (b) High-resolution d$I$/d$V$ spectra recorded in the locations highlighted by the colored dots in (a), which correspond to the As- and U-terminated surfaces, respectively. Acquisition parameters: (a) $V_{\mathrm{s}} = 1$ V, $I_{\mathrm{t}} = 15$ pA, $T$ = 4.2 K. (b) $V_{\mathrm{a.c.}} = 0.3$ mV, $T$ = 4.2 K.}
    \label{Figure S3:STM}
\end{figure*}

Fig.~\ref{Figure S3:STM}a shows a large-scale STM topography image of the UAsS sample where two atomic terraces can be observed. One of them (right) is the already discussed U-terminated surface while the region on the left corresponds to the As-terminated surface (see \ref{Figure S1:STM} for more details). To characterize and to establish a trustworthy comparison regarding the low-energy quasiparticle excitation near $E_{F}$ on the As-termination, high-resolution differential tunneling conductance spectra were recorded simultaneously with the very same tip-apex on the locations indicated by the color dots in Fig.~\ref{Figure S3:STM}a. The results, shown in Fig.~\ref{Figure S3:STM}b, point towards the absence of ZBA on the As-termination. This utterly demonstrates that the ZBA is not a bulk phenomenon but rather a surface related excitation peak that is probably inherent to the U-terminated surfaces.

\subsubsection{Influence of perpendicular magnetic field on the ZBA}\label{subsubsec:influence}

\begin{figure*}
    \centering
    \includegraphics[width=0.70\linewidth]{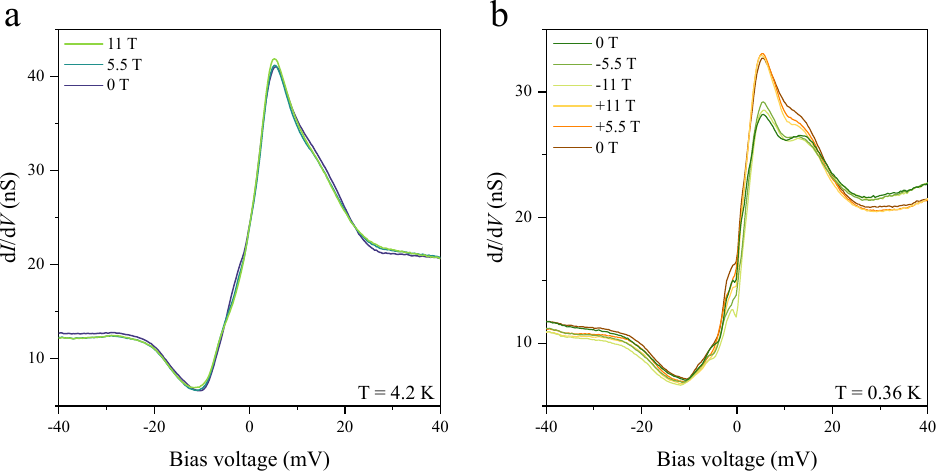}
    \caption{(a) Perpendicular magnetic field ($H_{\perp}$) dependence of the ZBA measured at 4.2 K. (b) Evolution of the ZBA under the presence of varying magnetic field direction. Note that between positive and negative fields values, there was a slight tip change. All data were spatially averaged within the same region of 3$\times$3 nm$^{\mathrm{2}}$. Acquisition parameters: (a) $V_{\mathrm{a.c.}} = 0.3$ mV, $T$ = 4.2 K; (b) $V_{\mathrm{a.c.}} = 0.3$ mV, $T$ = 0.36 K.}
    \label{Figure S4:STM}
\end{figure*}

We have demonstrated in the main text that the observed ZBA is not sensitive to the application of magnetic fields perpendicular to the sample surfaces, showing a strong resilience to fields strength up to 11 T at 0.36 K. This robustness against magnetic field perturbation also holds for data recorded at 4.2 K (Fig.~\ref{Figure S4:STM}a), as well as for varying fields direction (Fig.~\ref{Figure S4:STM}b).

\subsubsection{Temperature-dependent evolution of the ZBA and Fano line fitting analysis}\label{subsubsec:temperature}

\begin{figure*}
    \centering
    \includegraphics[width=0.70\linewidth]{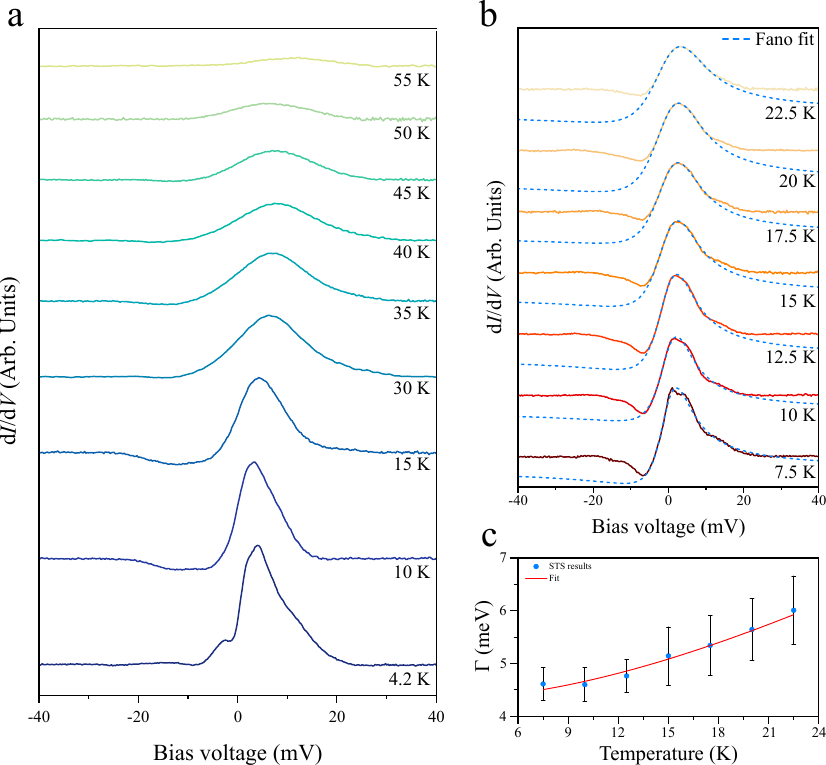}
    \caption{(a) Extended data set tracking the temperature dependence evolution of the ZBA in UAsS up to 55 K. (b) Additional data set showing the temperature evolution of ZBA measured on the same sample location (solid lines). The dashed blue lines show the corresponding thermally convoluted Fano line fit of the experimental data. (c) Extracted values of the intrinsic width, $\Gamma$, of the ZBA from each Fano fits shown in (b). Red solid line indicates the expected $\Gamma$($T$) evolution of Kondo resonance based on the Fermi liquid theory. Acquisition parameters: (a)-(b) $V_{\mathrm{a.c.}} = 0.3$ mV}
    \label{Figure S5:STM}
\end{figure*}

We have shown in the main text that the width of the ZBA peak broadens with temperature until ultimately vanishes upon reaching a certain treasure value. This behavior can be clearly observed in Fig.~\ref{Figure S5:STM}a for a more extended temperature range. This evolution with temperature is compatible with the expected characteristic of a Kondo resonance peak, where the intrinsic Kondo width should gradually broaden due to electron-electron scattering with increasing temperature according to the Fermi liquid theory\cite{Nagaoka2002,Kruger2005,Ternes2008,Gruber2018}.

To extract the intrinsic Kondo width ($\Gamma$) evolution with temperature from the experimentally measured data, we have carried out a thermally convoluted Fano line shape fit using the following equation:
\begin{equation}\label{eq: KondoSTSFit}
    \frac{dI}{dV}(eV) \propto \int_{-\infty}^{+\infty} \frac{df(\epsilon)}{d\epsilon}\vert_{\epsilon=E-eV}(a\frac{(\omega+q)^2}{1+\omega^2}+c)dE\, ; \quad \omega = \frac{E-E_0}{\Gamma}
\end{equation}
where $f$($\epsilon$) is the Fermi-Dirac distribution that accounts for the temperature broadening effect, and the Fano line shape is captured by the second term in the integrand, which represents a scaled Fano function plus a constant background. Here, $q$ represents the Fano line asymmetry, and $E_0$ is resonance energy position. During the fitting procedure, $E_0$, $q$, $\Gamma$, $a$, and $c$ are set as free fitting parameters at each fixed temperature. This formula captures well the experimental data as can be noticed in Fig.~\ref{fig:STM}f and also Fig.~\ref{Figure S5:STM}b, thus allowing us to extract the $\Gamma$($T$) evolution dependence (see Fig.~\ref{fig:STM}g and Fig.~\ref{Figure S5:STM}c)

\section{Details on the mean-field theory} \label{MFT}
In this section, we provide further details on our mean-field theory (MFT) calculation and the numerical procedure through which the MFT parameters were obtained. As a starting point, we consider the Kondo-Heisenberg Hamiltonian 
\begin{eqnarray}
\mathcal{H}&=&\mathcal{H}_{c}+\mathcal{H}_K+\mathcal{H}_{\rm FM}\\
    \mathcal{H}_{c}&=& -t^{(1)} \sum_{\langle ij \rangle\sigma}(c_{i\sigma}^\dagger c_{j\sigma}+c_{j\sigma}^\dagger c_{i\sigma}),\nonumber\\
    \mathcal{H}_{K}&=& \frac{J_K}{2}\sum_{i,\alpha,\beta} c_{i\alpha}^\dagger \vec{\sigma}_{\alpha\beta}c_{i\beta} \cdot \vec{S}_i\nonumber,\\
    \mathcal{H}_{\rm FM}&=& -J\sum_{\langle i j \rangle}\vec{S}_i \cdot \vec{S}_j,
\end{eqnarray}
where $i,j$ label the lattice sites, and $\sigma$ the spin-$1/2$ values. In addition, we consider the above Hamiltonian on a square lattice defined by a ribbon of dimensions $N_x\times N_y$, where $N_y$ corresponds to the number of sites along the periodic boundary direction, and $N_x$ to the remaining direction where an open boundary is considered.
We proceed to map the spin degrees of freedom into fermionic degrees of freedom using slave-fermions,
\begin{eqnarray}
    S_{i}^z&=&\frac{1}{2}\left(f^\dagger_{i\uparrow}f_{i\uparrow}-f^\dagger_{i\downarrow}f_{i\downarrow}\right)\nonumber\\
S_{i}^+&=&f^\dagger_{i\uparrow}f_{i\downarrow}\nonumber\\
    S_{i}^-&=&f^\dagger_{i\downarrow}f_{i\uparrow}\nonumber,
\end{eqnarray}
with the constraint that there must be $1$ $f$-fermion per site. This constraint can be fulfilled on average by the introduction of a Lagrange multiplier $\lambda$. Using the slave-fermion approach, we produce a mean-field decoupling into the hybridization and magnetization channels for the Kondo term, and the magnetization and particle-hole channel for the Heisenberg term. The resulting Hamiltonian is provided by the expression
\begin{eqnarray}
    \mathcal{H}^{\rm MFT}&=&\sum_{q} \psi_{q,\sigma}^\dagger\mathbb{H}_q^\sigma \psi_{q,\sigma} +J_K \sum_i  r_i^\uparrow r_i^\downarrow-J_K\sum_{i}m_{i}^f m_{i}^c-J\sum_{\langle ij \rangle}(\chi_{ij}^{\downarrow}\chi_{ij}^{\uparrow}-m_{i}^fm_{j}^f) -N\lambda,\nonumber\\
    &=&\sum_{q} \psi_{q,\sigma}^\dagger\mathbb{H}_q^\sigma \psi_{q,\sigma} +\mathcal{E}_0,
\end{eqnarray}
where 
\begin{eqnarray}
    \psi_{q,\sigma}^\dagger&=&\begin{pmatrix}
        c^\dagger_{0,q,\sigma}\ ,\ c^\dagger_{1,q,\sigma}\ ,\ c^\dagger_{2,q,\sigma}\ ,\  \cdots \ ,\ c^\dagger_{L_x-1,q,\sigma}, f^\dagger_{0,q,\sigma}\ ,\ f^\dagger_{1,q,\sigma}\ ,\ f^\dagger_{2,q,\sigma}\ ,\  \cdots \ ,\ f^\dagger_{N_x-1,q,\sigma}
    \end{pmatrix},\nonumber\\
    \mathbb{H}^\sigma_q&=&\begin{pmatrix}
        \mathbb{H}_{q,cc}^{\sigma} & \mathbb{H}_{q,cf}^{\sigma} \\
        \mathbb{H}_{q,fc}^{\sigma} & \mathbb{H}_{q,ff}^{\sigma}
    \end{pmatrix},
\end{eqnarray}
with the blocks of this matrix $4N_x \times 4N_x $ matrix being defined as
{\small
\begin{eqnarray}
\mathbb{H}_{q,cc}^\sigma&=&\begin{pmatrix}
            -\mu +\sigma\frac{J_K}{2}m^f_0  -2t\varepsilon(q) & -t & 0 &  \cdots\\
            -t & -\mu+\sigma\frac{J_K}{2}m^f_1 -2t\varepsilon(q) & -t&\cdots\\
            0 &-t & -\mu+\sigma\frac{J_K}{2}m^f_2 -2t\varepsilon(q) & \ddots \\
            0 & 0 &  \ddots & \ddots
        \end{pmatrix}, \nonumber\\
\mathbb{H}_{q,ff}^\sigma&=&\begin{pmatrix}
            \lambda +\frac{\sigma}{2}(J_K m^c_0-J[2m^f_0+m^f_1]) +J\chi_0^{\bar{\sigma}}\varepsilon(q) & \frac{J}{4}\chi_{(0,1)}^{\bar{\sigma}} & 0 & \cdots\\
            \frac{J}{4}\chi_{(0,1)}^{\bar{\sigma}} & \lambda+\frac{\sigma}{2}(J_K m^c_1-J[m^f_0+2m_1^f+m^f_2]) + J\chi_1^{\bar{\sigma}}\varepsilon(q)&\frac{J}{4}\chi_{(1,2)}^{\bar{\sigma}}&\cdots\\
            0 & \frac{J}{4}\chi_{(1,2)}^{\bar{\sigma}}  & \ddots& \cdots \\
              \vdots & \vdots &  \vdots &  \ddots
        \end{pmatrix},\nonumber\\
        \mathbb{H}_{q,cf}^\sigma&=&\begin{pmatrix}
             -\frac{J_K}{2}r^{\bar{\sigma}}_0 &0 & 0 &  \cdots\\
            0 &  -\frac{J_K}{2}r^{\bar{\sigma}}_1 & 0&\cdots\\
            0 & 0 &  -\frac{J_K}{2}r^{\bar{\sigma}}_2 & \ddots \\
            0 & 0 &  \ddots & \ddots\nonumber
        \end{pmatrix},\label{eq:final_MFT_matrix}
\end{eqnarray}
}
where $\varepsilon(q)=\cos(q)$, and  the $c$-electrons and $f$-fermions are now characterized by three labels: the $x$ position in the ribbon, the $q$ momentum along the periodic direction, and the spin index $\sigma$. In the above equations, the mean-field parameters take the form
\begin{eqnarray}
    r_x^\sigma&=& \frac{1}{2N_y}\sum_{q}\langle f^\dagger_{(x,q),\sigma}c_{(x,q),\sigma}+ c^\dagger_{(x,q),\sigma}f_{(x,q),\sigma}\rangle\nonumber \\
    m_x^c&=& \frac{1}{2N_y}\sum_{q,\sigma}\sigma\langle c^\dagger_{(x,q),\sigma}c_{(x,q),\sigma}\rangle\nonumber \\
     m_x^f&=& \frac{1}{2N_y}\sum_{q,\sigma}\sigma\langle f^\dagger_{(x,q),\sigma}f_{(x,q),\sigma}\rangle\nonumber \\
    \chi_{(x,x+\delta_x)}^\sigma&=&\frac{1}{2N_y}\sum_{q}\langle f^\dagger_{(x,q),\sigma} f_{(x+\delta_x,q),\sigma}+ f^\dagger_{(x+\delta x,q),\sigma} f_{(x,q),\sigma}\rangle\nonumber\\
      \chi_{(x)}^\sigma&=&\frac{1}{N_y}\sum_{q} \cos(q) \langle f^\dagger_{(x,q),\sigma} f_{(x,q),\sigma}
      \rangle\nonumber\\
      \lambda&\to& \frac{1}{N_y}\sum_{x,q,\sigma}\langle f^\dagger_{(x,q),\sigma}f_{(x,q),\sigma}\rangle=1.
\end{eqnarray}
Using the above mean-field Hamiltonian and its corresponding mean-field self-consistency equations, we proceed to numerically determine a set of MFT parameters that minimize the free-energy 

\begin{eqnarray}
    \mathcal{F}    &=&\mathcal{E}_0 -\frac{1}{\beta}\sum_{(q,x)}\ln\left(1+e^{-\beta \varepsilon_{x}(q)}\right),
\end{eqnarray}
where $\beta$ is the inverse temperature and $\varepsilon_x(q)$ labels the corresponding the eigenvalues. In addition to minimizing the free-energy, we also require the MFT parameters to preserve a fixed filling of 1 $f$-fermion per site at various $c$-electron fillings $n_c$. 

For the self-consistent calculation, we have considered the hopping coupling $t$ as the measure of energy, studied systems with a small and high Heisenberg coupling $J$, i.e. $J\in \{ 0.1,2 \}$, and vary the Kondo coupling $J_K$. For a particular geometry $N_x\times N_y$, a total of $8N_x$ position-dependent MFT parameters fulfilling $8N_x$ self-consistency equations must be determined. Our self-consistent calculations are carried out using a bootstrap approach at a temperature $T=t\times 10^{-4}$ for $n_c\in \{0.5,1\}$. In this calculation, convergence of the MFT parameters is considered once the average collective change of the MFT parameters is below certain threshold which we typically set to be $t \times 10^{-6}$. To find the optimal MFT parameters we determined up to 250 distinct $x$-dependent MFT sets per set of interaction parameters $(t,J,J_K)$, and selected the one for which the free-energy is minimal.

\begin{figure}[ht!]
    \centering
    \includegraphics[width=\linewidth]{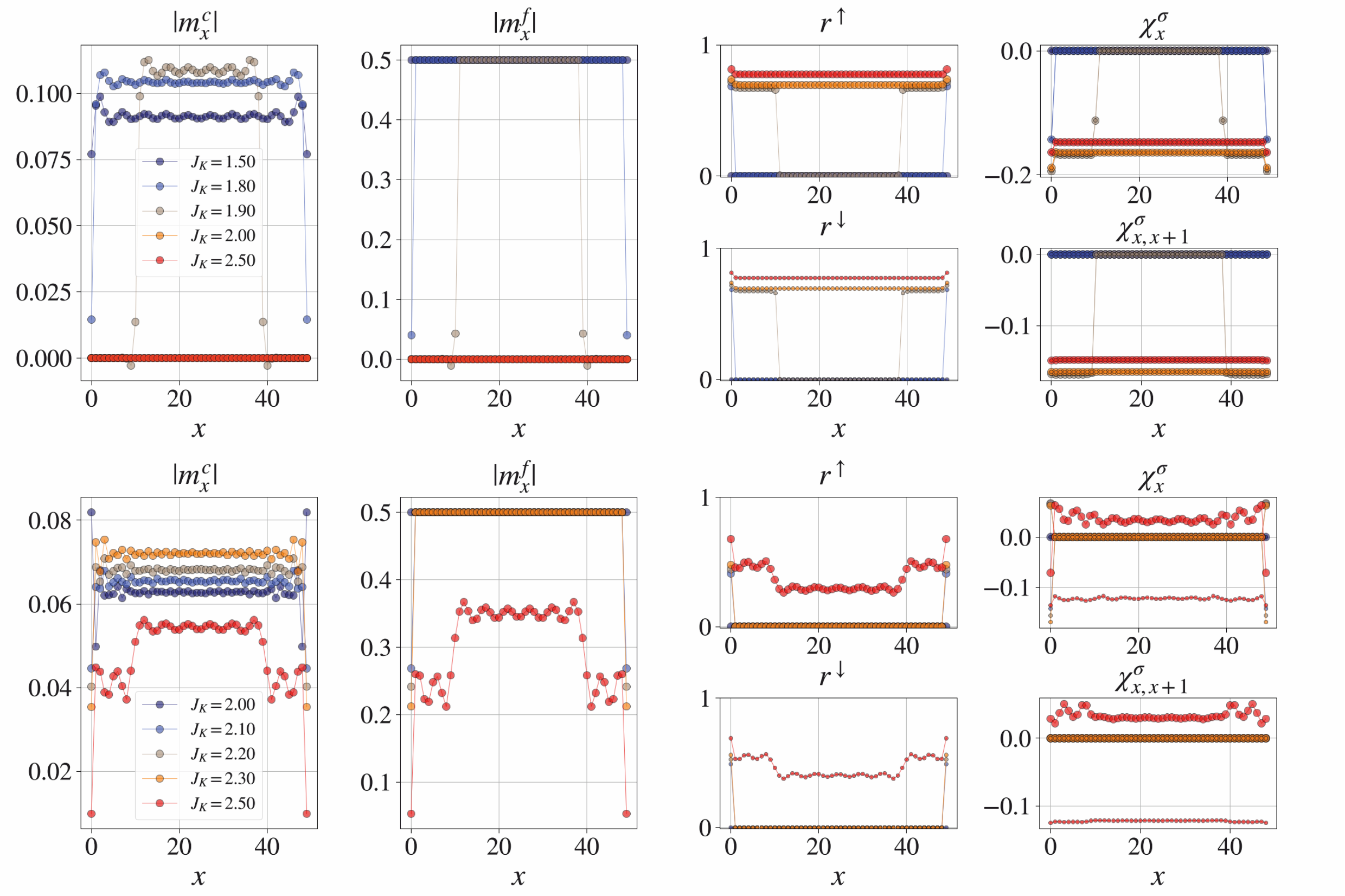}
    \caption{Spatial dependence of the mean-field parameters for the Heisenberg-Kondo model with $J=0.1t$ for distinct values of $J_K$ at half-filling (upper row) and at quarter-filling (lower row).}
    \label{fig:MFT_params_J0.1}
\end{figure}

\begin{figure}[ht!]
    \centering
    \includegraphics[width=\linewidth]{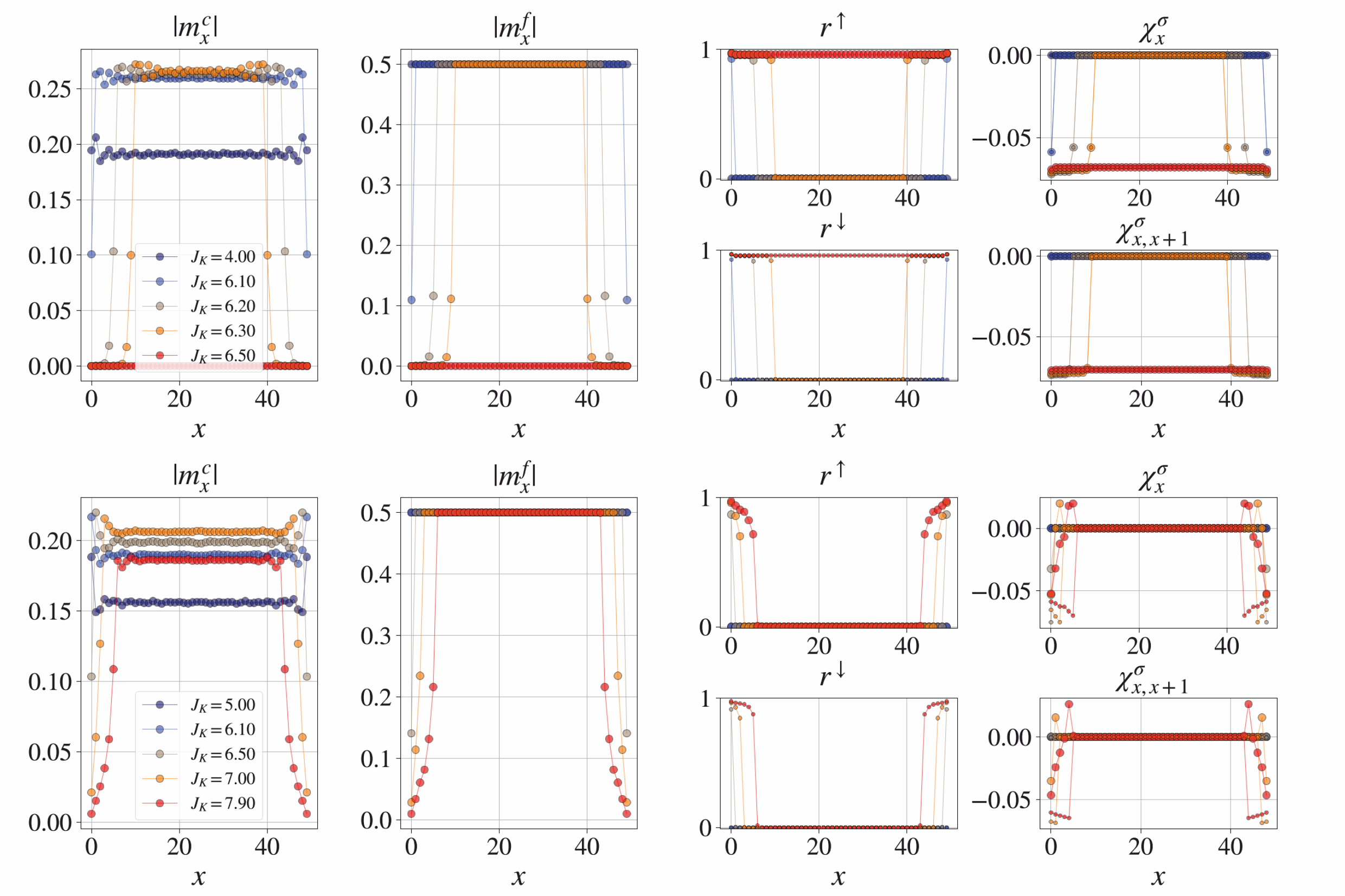}
    \caption{Spatial dependence of the mean-field parameters for the Heisenberg-Kondo model with $J=2t$ for distinct values of $J_K$ at half-filling (upper row) and at quarter-filling (lower row).}
    \label{fig:MFT_params}
\end{figure}



As discussed in the main text, the physics of the Heisenberg-Kondo model is governed by the interplay between the Heisenberg coupling and the Kondo coupling; at low Kondo coupling, the system is expected to show a vanishing hybridization as well as a non-zero $c$-electron and $f$-fermion magnetization. On the other hand, for a strong $J_K$ coupling a Kondo insulator is expected. Figure~\ref{fig:MFT_params} illustrates the $x$-dependence of the MFT parameters for a system with $J=2t$ at various Kondo coupling strengths at half- (upper row) and quarter-filling (lower row), i.e. $n_c=1$ and $n_c=0.5$, respectively. In the following discussion we mainly focus on the results for half-filling. As shown in the upper row of Fig.~\ref{fig:MFT_params}, for the smallest Kondo coupling illustrated, only the magnetization MFT parameters are non-vanishing. The oscillatory character of the magnetization parameters is produced by the introduction of the boundary, i.e. these become uniform when the boundary is absent. In contrast, for the highest Kondo coupling, the magnetization parameters vanish, while the hybridization approaches its saturation value of $1$ (it is worthwhile noting that, in this limit, the hybridization is slightly stronger in the boundary than in the bulk). Interestingly, for intermediate Kondo couplings $J_K$ an on-site hybridization $r_x^\sigma$, signaling the appearance of heavy-fermions, is observed at the boundaries of the system along with a diminished magnetization of both the $c$-electrons and the $f$-fermions, while, for sites away from the boundary, no hybridization is present. In other words, the electrons in the vicinity of the boundary start to localize forming Kondo singlets, while those in the bulk remain conductive. Such effect implies that the introduction of a boundary to the system manufactures an intermediate phase where the a Heavy-fermion FM phase first develops at the boundary and slowly starts invading the bulk as the Kondo coupling is increased, see the progression of the hybridization parameter as a function of site $x$ and Kondo couplings $J_K$. In practical terms, the appearance of this boundary phase implies that a bulk prove would indicate a metallic behavior, while a surface prove would hint towards an insulating character. 

\begin{figure}[ht!]
    \centering
    \includegraphics[width=\linewidth]{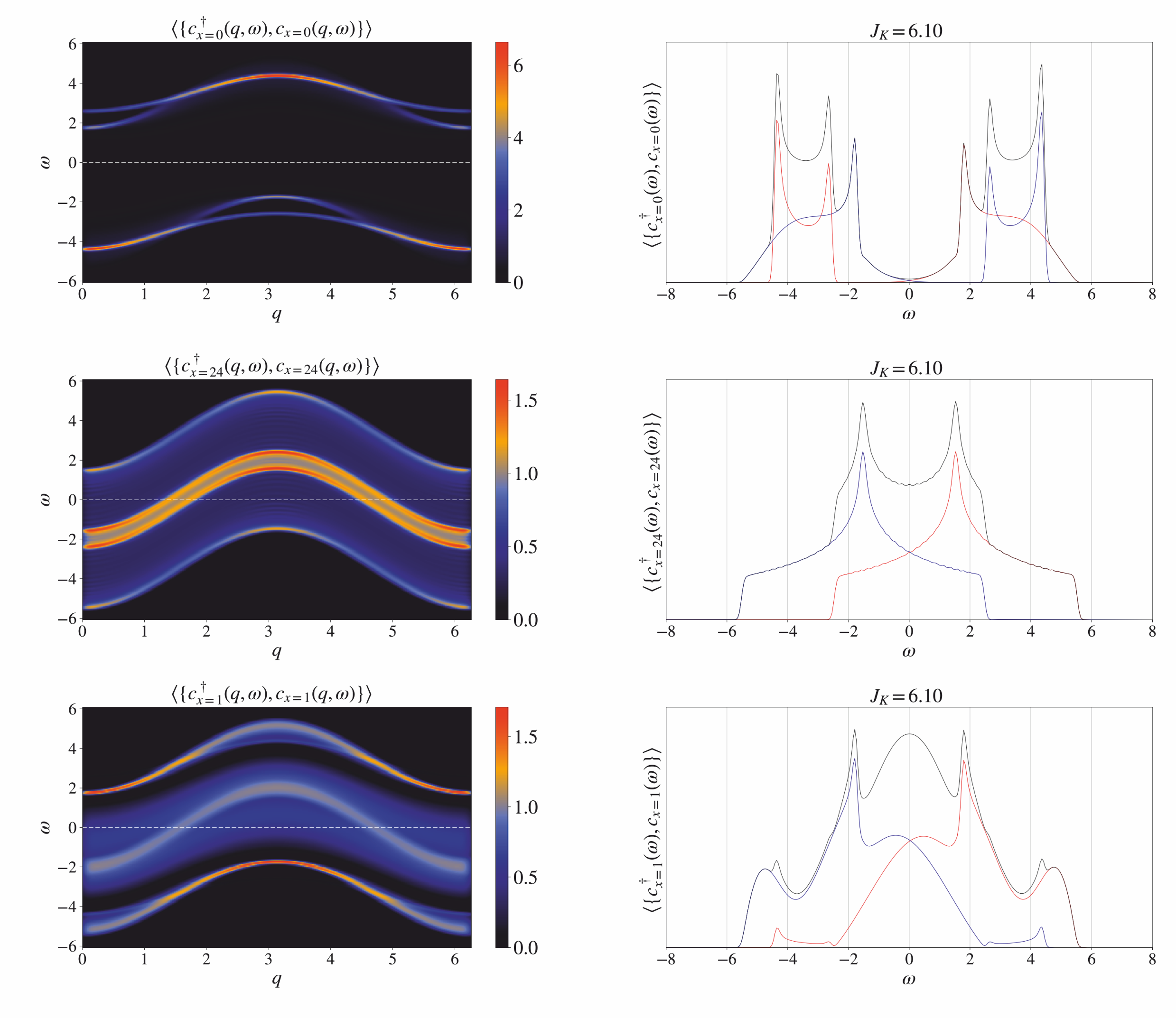}
    \caption{$q$ and $\omega$ (left column), and $q$-integrated (right column) spatial-dependent spectral function at the boundary, i.e. $x=0$ (upper row), at the first site away from the boundary, i.e. $x=1$ (middle row), and deep into the bulk, i.e. $x=24$ (lower row) for a system with $(J,J_K)=(2,6.1)$ at half filling, i.e. $n_c=1$. On the left column, the red, blue and black curves correspond to the contribution from the $\uparrow$ spins, the $\downarrow$ spins, and the combined contribution.}
    \label{fig:Bands_spectral_J2_nc1}
\end{figure}

To further study the implications of the intermediate boundary phase we study the space-resolved spectral function of the $c$-electrons and the $f$-fermions defined as
\begin{eqnarray}
    A_{cc}^\sigma(x,\omega)&\equiv&\sum_q \langle\{ c^\dagger_{(x,q),\sigma}(\omega),c_{(x,q),\sigma}(\omega)\}\rangle\\
    A_{ff}^\sigma(x,\omega)&\equiv&\sum_q \langle\{ f^\dagger_{(x,q),\sigma}(\omega),f_{(x,q),\sigma}(\omega)\}\rangle,
\end{eqnarray}
where $\{\cdot,\cdot\}$ is the anti-commutator. We now compute the spatial-dependent spectral function at $J_K=6.2$ for two distinct sites, one at the boundary and another deep into the bulk, see left column of Fig.~\ref{fig:Bands_spectral_J2_nc1}. As shown in Fig.~\ref{fig:Bands_spectral_J2_nc1}, for $x=0$ a set of insulating hybridized heavy-fermion bands yield the highest intensity for the spectral function, reflecting the formation of the Kondo-singlets at the boundary of the system. Interestingly, if we now consider the spectral function for the first site away from the boundary (i.e. $x=1$), the weight of the spectral function is distributed between a conductive continuum set of bands around $\omega=0$ (steaming from the bulk) and  a separate contribution coming from the hybridized bands above and below said continuum.  In contrast, for the site deep into the bulk ($x=24$) a broad continuum of bands present the highest intensity, reflecting the conductive nature of the bulk. The conductive or insulating nature of these spectral functions can be further expose by considering the $q$-integrated spectral function shown in the right column of Fig.~\ref{fig:Bands_spectral_J2_nc1}. Indeed, for the site at the boundary a diminished intensity is observed at the Fermi energy, in addition to the observation of higher-intensity peaks associated with the hybridized bands. For the first site away from the boundary, the spectral function reveals a set of sharp features (at $\omega\simeq \pm 2$) associated with the hybridized modes, and a broad continuum steaming from the conductive bulk. Lastly, for the site deep into the bulk, the typical DOS associated with the square lattice is exposed, where the two peaks come from the distinct Zeeman-split spin sectors of the system.

